\documentclass[twoside]{IEEEtran}
\usepackage{caption}
\usepackage{subcaption}
\usepackage{cite}
\usepackage{fancyhdr}
\usepackage{graphics} 
\usepackage{epsfig} 
\usepackage{epstopdf}
\usepackage{amssymb}
\usepackage{extarrows}
\usepackage{balance}
\usepackage{multirow}
\usepackage{float}
\usepackage[latin1]{inputenc}
\usepackage{setspace}
\usepackage{commath}
\usepackage{color,soul}
\usepackage[table]{xcolor}
\usepackage{hyperref}
\ifCLASSINFOpdf
\else
\fi
\begin{document}
\bibliographystyle{IEEEtran}
%
\title{Sensing Data Fusion for Enhanced Indoor Air Quality Monitoring}
%
%
%
\author{ Q. P. Ha, S. Metia, and M. D. Phung
\thanks{Q. P. Ha and S. Metia are with Faculty of Engineering and Information Technology, University of Technology Sydney, Broadway NSW 2007, Sydney, Australia (e-mail: Quang.Ha@uts.edu.au, Santanu.Metia@uts.edu.au, and ManhDuong.Phung@uts.edu.au).}
\thanks{M. D. Phung is with Faculty of Engineering and Information Technology, University of Technology Sydney, Broadway NSW 2007, Sydney, Australia and also with the Department of Electronics
	and Computer Engineering, University of Engineering and Technology,
	Vietnam National University, 144 Xuan Thuy, Cau Giay Hanoi 100000,
	Vietnam (e-mail: duongpm@vnu.edu.vn)}}

\maketitle

\begin{abstract}
Multisensor fusion of air pollutant data in smart buildings remains an important input to address the well-being and comfort perceived by their inhabitants. An integrated sensing system is part of a smart building where real-time indoor air quality data are monitored round the clock using sensors and operating in the Internet-of-Things (IoT) environment. In this work, we propose an air quality management system merging indoor air quality index (IAQI) and humidex into an enhanced indoor air quality index (EIAQI) by using sensor data on a real-time basis. Here, indoor air pollutant levels are measured by a network of waspmote sensors while IAQI and humidex data are fused together using an extended fractional-order Kalman filter (EFKF). According to the obtained EIAQI, overall air quality alerts are provided in a timely fashion for accurate prediction with enhanced performance against measurement noise and nonlinearity. The estimation scheme is implemented by using the fractional-order modeling and control (FOMCON) toolbox. A case study is analysed to prove the effectiveness and validity of the proposed approach.
\end{abstract}


%
\IEEEpeerreviewmaketitle

\section{Introduction}
%
%
%
%

\IEEEPARstart{W}{}ith the increasing growth worldwide of active population working inside a building, the management of indoor air quality is becoming crucially important for human health and work efficiency \cite{Jaimini_2017}.  In this regard, the development of smart buildings is aimed to provide comfort and improved indoor air quality (IAQ) for occupants. Common issues associated with IAQ include improper or inadequately-maintained heating and ventilation as well as pollution by hazardous materials \cite{Lay-Ekuakille_2011} (olefins, aromatics, hydrocarbons, glues, fiberglass, particle boards, paints, etc.) and
other contaminant sources (laser printers \cite{Tang_2012}, tobacco smoke, excessive concentrations of bacteria, viruses, fungi (including molds \cite{Sahin_2016}), etc.). Moreover, the increase in the number of building occupants and the time spent indoors directly impact the IAQ \cite{Jin_2018}. 

Air quality can be evaluated by such parameters as concentration
of air pollutants including carbon monoxide (CO), carbon dioxide (CO$_2$), formaldehyde (HCHO), nitrogen dioxide (NO$_2$), ozone (O$_3$), sulfur dioxide (SO$_2$), total volatile organic compounds (TVOCs), particulate matter (PM$_{2.5}$), total suspended particles (TSP), as well as temperature, relative humidity and air movement. For an indoor environment, air quality is affected also by household chemicals, furnishings,  air contaminants emitted from outside, occupant activities (e.g., smoking, cooking, breathing) \cite{Zimmermann_2018}, as well as air infiltration, and manual/mechanical ventilation. Due to a large number of factors involved,  the development of an accurate system for IAQ monitoring is of great interest. To this end, fusing heterogeneous data from a network with a multitude of sensor types is essential for calculating the IAQI and for monitoring different pollutants in a building. 

	For indoor air quality, season-dependent models have been developed in \cite{Kim_2012} for monitoring, prediction and control of the IAQ in underground subway stations, where the IAQ of a metro station is shown to be influenced by temperature variations in different seasons. As indoor air quality (IAQ) is affected by heating, ventilation and air-conditioning conditions, modelling and control strategies have been proposed for residential air conditioning \cite{Vakiloroaya_2013} and ventilation systems \cite{Sun_2013} to improve the occupants' living environment. A recent work \cite{Zhao_2018} has showed that IAQ is affected by both outdoor particle concentration and indoor activities (walking, cooking, etc.). In  \cite{Merabtine_2018}, IAQ is assessed by monitoring and analysing CO$_2$ levels at the building's foyer area taking into account also thermal comfort. While indoor thermal comfort can be predicted via himidity \cite{Rana_2017}, it is known that an elevated level of humidity may have a positive impact on the perceived IAQ with some effects on human health \cite{Wolkoff_2018}. On the other hand, for IAQ modeling, data fusion is an effective way to reduce the sensors measurement uncertainties and overcome sensory limitations \cite{Assa_2015}. Various strategies have been employed, among which  Kalman filtering is quite popular and effective. Multi-sensor data fusion using Kalman filtering is adopted in \cite{Carminati_2012} to estimate the mass and flow parameters of gas transport processes from their relation to energy consumption and air quality in an indoor environment. For improving the model accuracy and robustness, system identification and data fusion are implemented for on-line adaptive energy forecasting in virtual and real commercial buildings with filter-based techniques \cite{Li_2016}. In \cite{Wang_2017}, a Kalman consensus filter is also used to analyze aircraft cabin contamination data with state estimation.

Motivated by \cite{Assa_2015, Rana_2017, Wolkoff_2018}, this paper proposes a data fusion strategy for the sensor network of a smart building to integrate the humidex and IAQI into an Enhanced Indoor Air Quality Index (EIAQI) with a weighting scheme to take into account also indoor humidity. Here, an Extended Fractional Kalman Filter (EFKF) incorporating the Mat\'{e}rn covariance function  and a fractional order system is developed to deal with spatial distributions as well as the highly nonlinear, uncertain nature of indoor air quality data while merging humidex into IAQI for the proposed EIAQI. Unlike existing works, here humidex is integrated into a proposed Enhanced Indoor Air Quality Index for IAQ assessment, and in terms of IAQ prediction in buildings, the proposed EFKF with a proper choice of the correlation length allows for improving accuracy of the air pollutants profiles in places where sensory measurements or data processing may overlook. This merit, for indoor environment, makes use of the advantage in using the Mat\'{e}rn covariance model to smooth the data obtained from sensing and to describe prominent nonstationary characteristics of the global environmental processes, where outdoor monitoring stations are too sparse for accurate assessment \cite{Jun_2014}. As such, the contributions of this paper rest with (i) the inclusion of humidity in assessing indoor air quality, and (ii) the ability to recover missing data collected from sensors with the use of Extended Fractional Kalman Filtering and Mat\'{e}rn function-based covariances to improve accuracy of IAQ prediction. 

{With the continuous development of pollutant monitoring sensor technology, atmospheric parameters monitoring sensor technology, IoT technology and  information communication technology (ICT), it is possible to  monitor IAQ real-time in which people work and live at any time \cite{Yue_2017,Zhao_2019}. In \cite{Chang_2019}, authors have  investigated how temperature affected the formaldehyde emission rate by wooden materials in a smart building using IoT sensors. In \cite{Liu_2019}, researchers have proposed and implemented a remote monitoring and management solution for a smart building. For monitoring and controlling of the indoor climate, authors have used a plant wall system based on the Azure public cloud platform which was based on IoT technology.}    

The remainder of the paper is organized as follows. Section II describes the sensor network system for IAQ management and the paper motivation. Section III presents the proposed framework for obtaining the enhanced indoor air quality index. In Section IV, the EFKF development is included together with results and discussion on real data obtained from the building network sensors. Rationale for data fusion with EFKF as well as IAQ assessment are given in Section V. Finally, the conclusion is drawn in Section VI.  

{A summary of major notations used in this paper is listed in Table~\ref{table_A}.}
\begin{table}[h!]
	\renewcommand{\arraystretch}{1.3}
	\caption{\textsc{ Brief Summary of Major Notations}}
	\label{table_A}
	\centering
	\footnotesize
	\scalebox{0.75}{	\begin{tabular}{|ll|ll|} 
		\hline
		$I_p$ &: Index for pollutant $p$ & $C_p$ &: Rounded concentration\\
		$BP_{ul}$ &: Breakpoint greater than $C_p$ & 	$BP_{ll}$ &: Breakpoint less than $C_p$\\
		$l_{ul}$ &: Index value corresponding $BP_{ul}$ & 	$l_{ll}$ &: Index value corresponding $BP_{ll}$\\ 
		$h$ &: Humidex & $T$ &: Temperature\\
		$RH$ &: Relative humidity & $IAQI$ &: Indoor air quality index\\
		$EIAQI$ &: Enhanced indoor air quality index & $W_h$ &: Humidex weighting factor\\
		$W_T$ &: Overall $EIAQI$ weightage & $l$ &: Correlation length\\
		$\lambda$ &: Positive constant & $\alpha$ &: Fractional order\\
		$k$ &: Discrete-time index & $\epsilon$ &: Output error\\ 
		$y_{r_i}$ &: Forecast value & 	$y_{r_i}$ &: Measured value\\
		$N$ &: Number of samples & $R^2$ &: Coefficient of determination\\
		\hline
	\end{tabular}}
\end{table}
      
\begin{figure}[t!]
	\centering
	\begin{minipage}[b]{0.22\textwidth}
		\includegraphics[width=\textwidth]{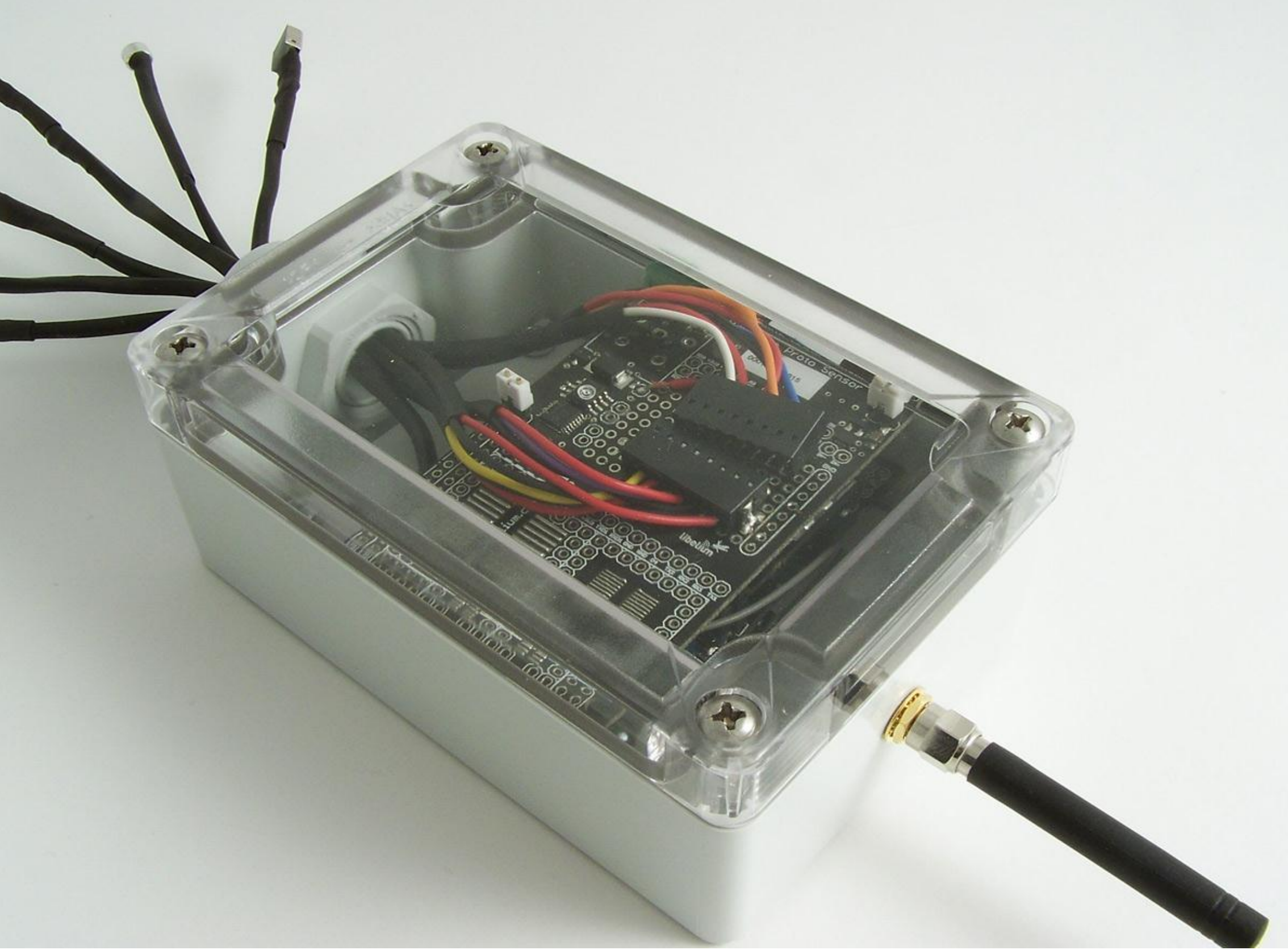}
		\caption{Waspmote sensor for recording IAQ.}
		\label{fig_1}
	\end{minipage}
	\hfill
	\begin{minipage}[b]{0.22\textwidth}
		\includegraphics[width=\textwidth]{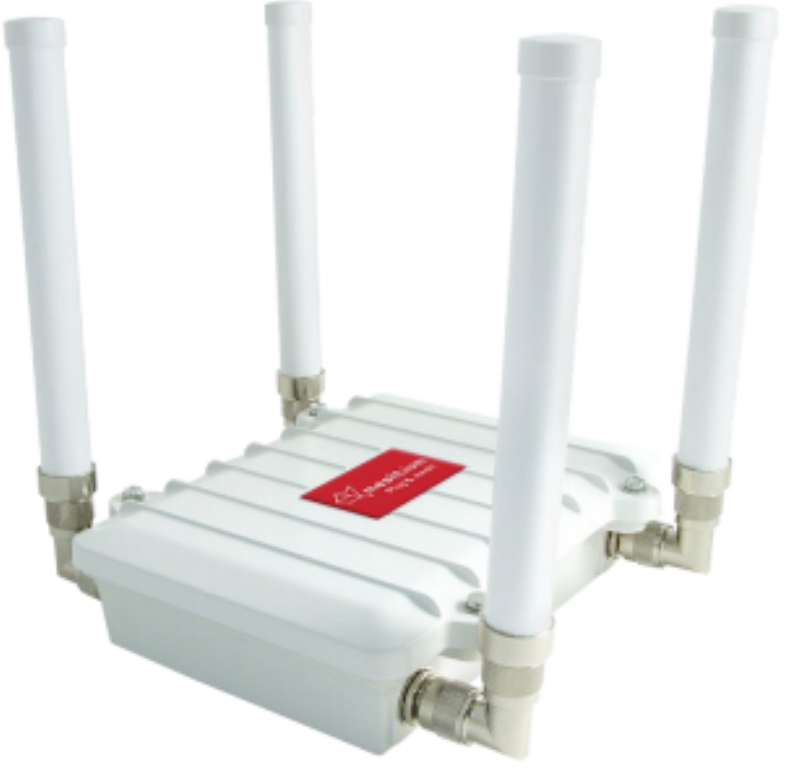}
		\caption{Meshlium gateway router}
		\label{fig_2}
	\end{minipage}
\end{figure}
\section{System Description and Motivation}
An office building is chosen as a testbed for the study in this paper. The building is embedded with numerous sensors for monitoring of its energy consumption as well as internal and external environment. For environmental monitoring, data are collected for such parameters as structural strain, people counting, vibrations and noise levels, as well as gas concentrations, weather, temperature, and meteorological conditions. 

\begin{figure*}[h!]
	\centering
	\includegraphics[width=0.95\textwidth]{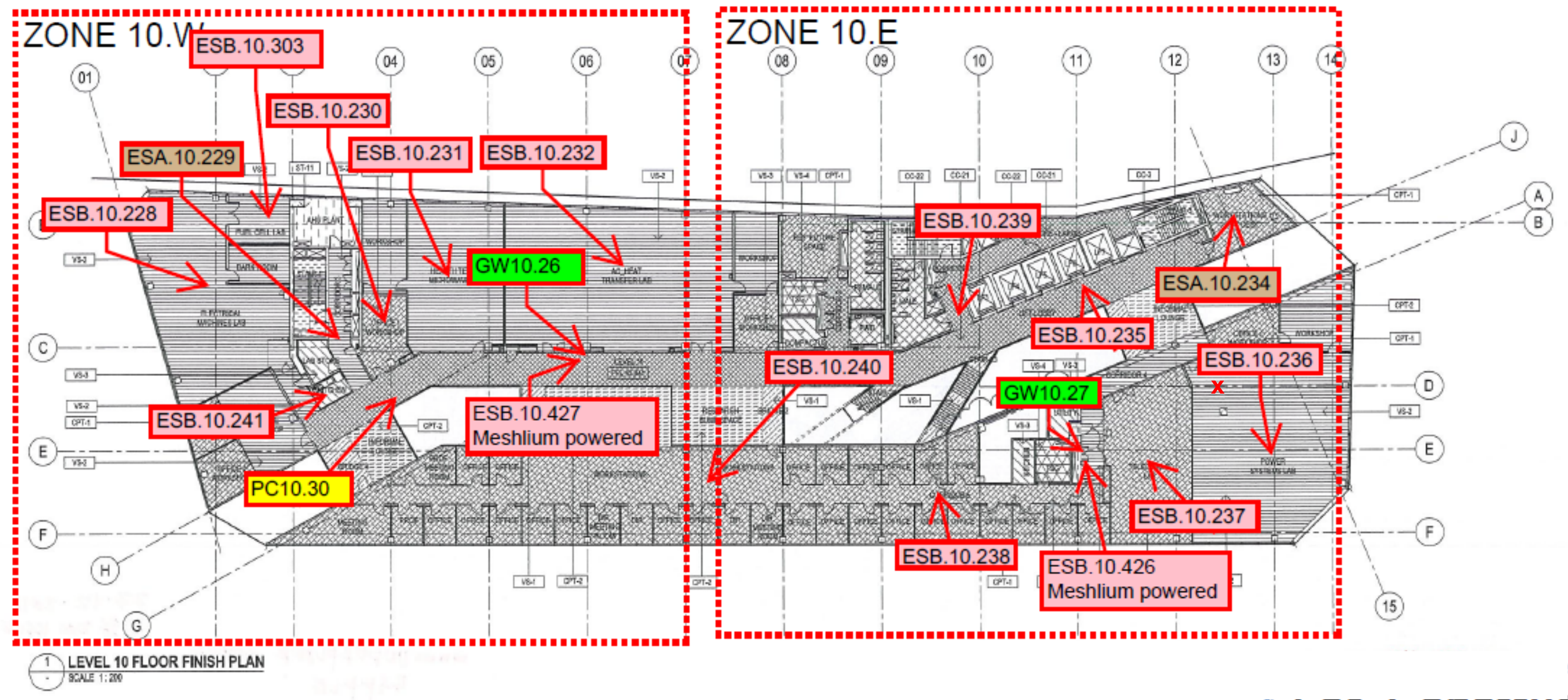}
	\caption{Floor plan and location of the waspmote sensors. }
	\label{fig_A}
\end{figure*}

The building management system is installed on its top floor. In this paper, our focus is on its application to the monitoring of IAQ in the building only. The sensors used in the building's IAQ sensor network are the waspmote sensors, as shown in Fig.~\ref{fig_1}. The sensor can measure levels of air pollutant (hydrogen (H$_2$), ammonia (NH$_3$), ethanol (C$_2$H$_6$O), hydrogen sulfide (H$_2$S), and toluene (C$_7$H$_8$)), as well as carbon monoxide (CO), carbon dioxide (CO$_2$) and oxygen (O$_2$) in parts per million (ppm). Temperature and humidity are also recorded in \textdegree C (centigrade) and \%RH (relative humidity). For air quality monitoring, 16 sensors were implemented on a floor and more than 100 others were implemented throughout the building. Sensor data gathered by the waspmote plug and sense nodes are sent to the cloud by the Meshlium, a gateway router specially designed to connect the waspmote sensor networks to the Internet via Ethernet,  Wi-Fi and 3G interfaces, as shown in Fig.~\ref{fig_2}. Location of the waspmote sensors (ESB.10.228 - ESB.10-427) installed on the west and east wings of the10$^{th}$ floor (10.W and 10.E) of the building of interest is shown in Fig.~\ref{fig_A}. 

It should be noted that sensitivity of waspmote sensors may vary from one unit to another in a wide range in dealing with various concentrations of different gases, such as H$_2$, NH$_3$, C$_2$H$_6$O, H$_2$S, C$_7$H$_8$, CO, CO$_2$ and O$_2$. Hence, a proper calibration procedure may be required subject to their operation range and conditions of the application to be implemented under controlled temperature and humidity. The larger the number of calibration points in that range the more accurate the monitoring. Moreover, it is also necessary to select suitable values load resistance and amplification gain for each waspmote sensor to adapt with its measurement range. As dependent on the way the sensor is supplied, the longer the power time or duty cycle, the better its accuracy. The tradeoff here is an increase in the mote's consumption, with the consequent decrease of the battery's life, which requires a regular check-up of the power supply. Moreover, the transport processes of emissions gases, in terms of mass and flow parameters at indoor temperature and humidity conditions, with respect to the energy consumption and building services may also affect the accuracy of waspmote sensors.
		
\begin{figure}[h!]
	\centering
	\includegraphics[width=0.45\textwidth]{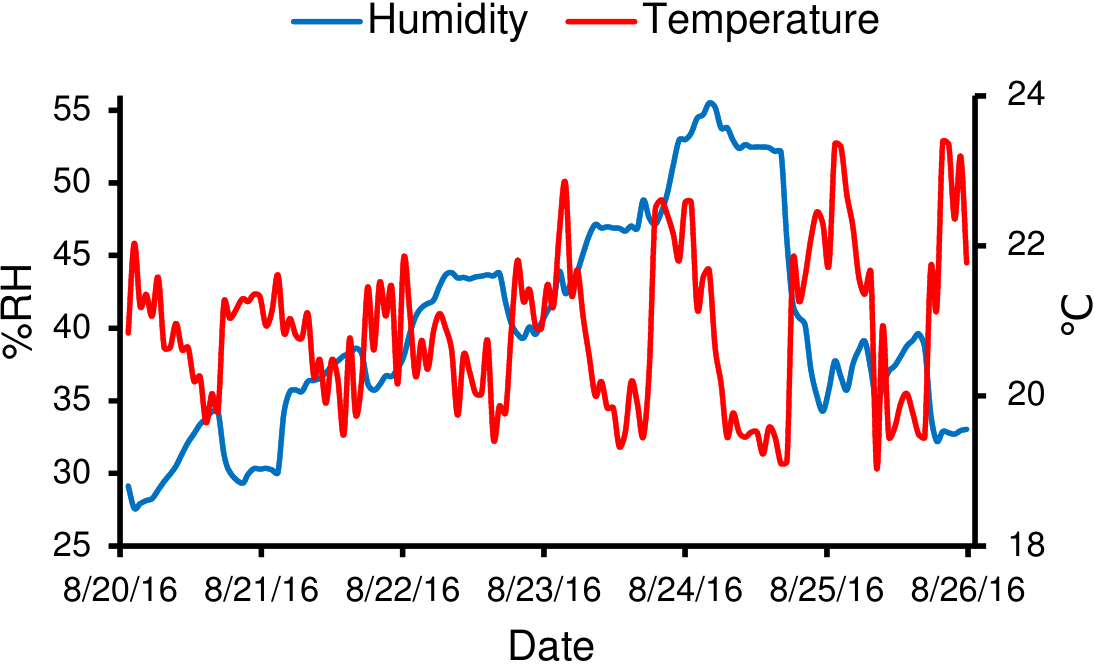}
	\caption{Temperature (\textdegree C) and humidity (\%RH) levels by waspmotes.}
	\label{fig_5}
\end{figure}

\begin{figure}[h!]
	\centering
	\includegraphics[width=0.45\textwidth]{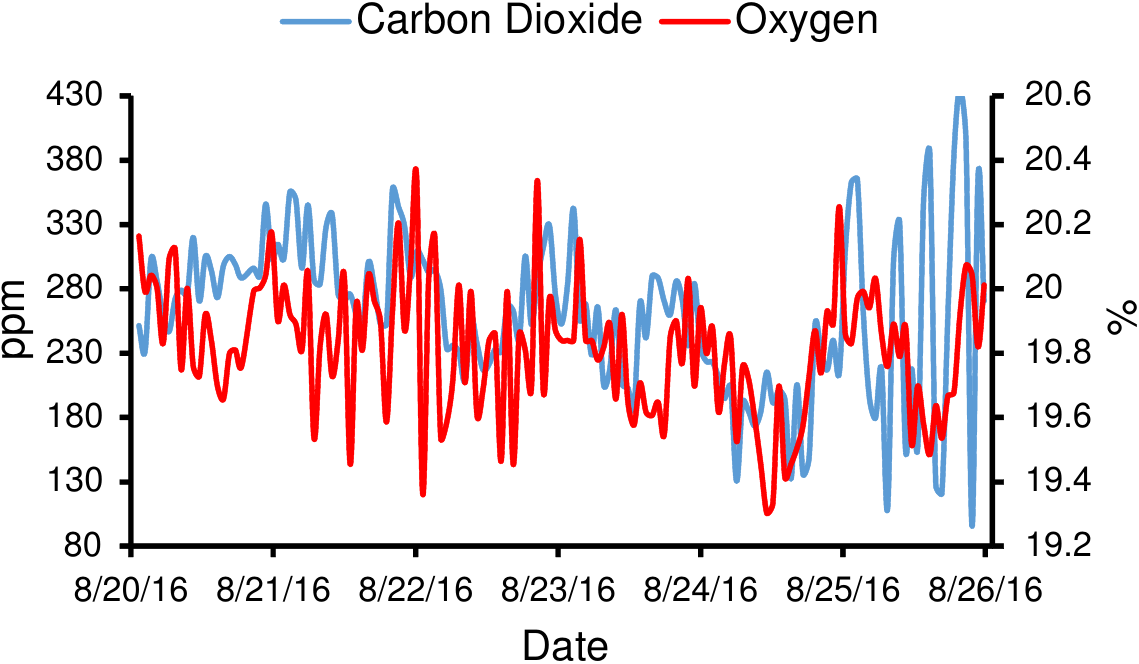}
	\caption{CO$_2$ (ppm)  and O$_2$ (\%) levels by waspmotes.}
	\label{fig_4}
\end{figure}

\begin{figure}[h!]
	\centering
	\includegraphics[width=0.45\textwidth]{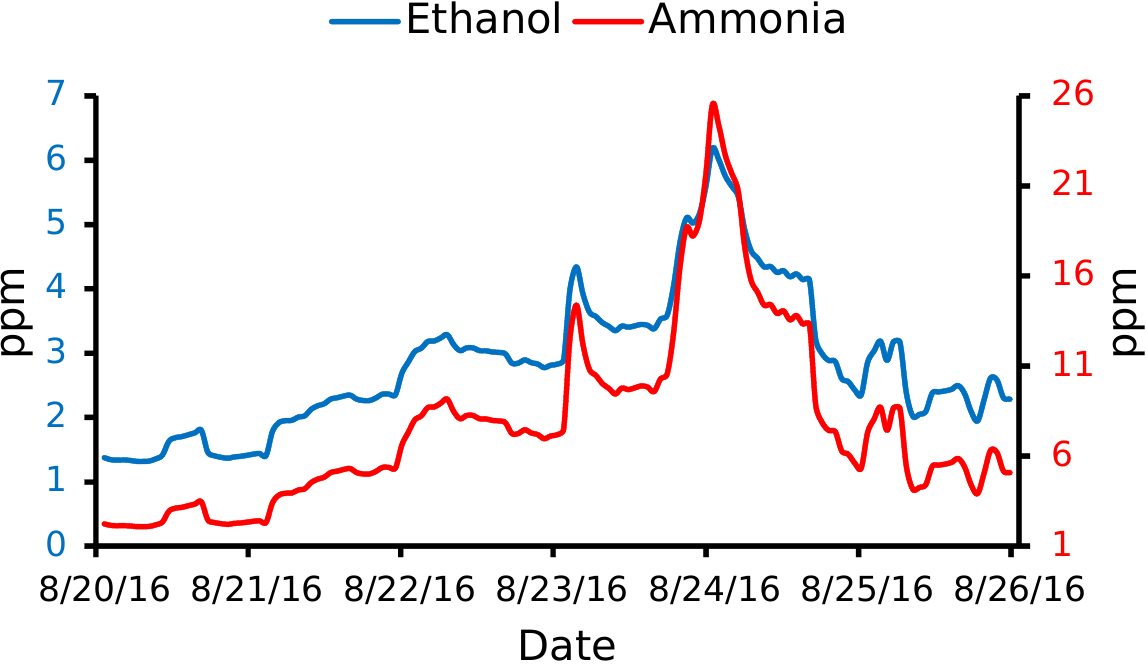}
	\caption{Ethanol (ppm) and Ammonia (ppm) levels by waspmotes.}
	\label{fig_4a}
\end{figure}

\begin{figure}[h!]
	\centering
	\includegraphics[width=0.45\textwidth]{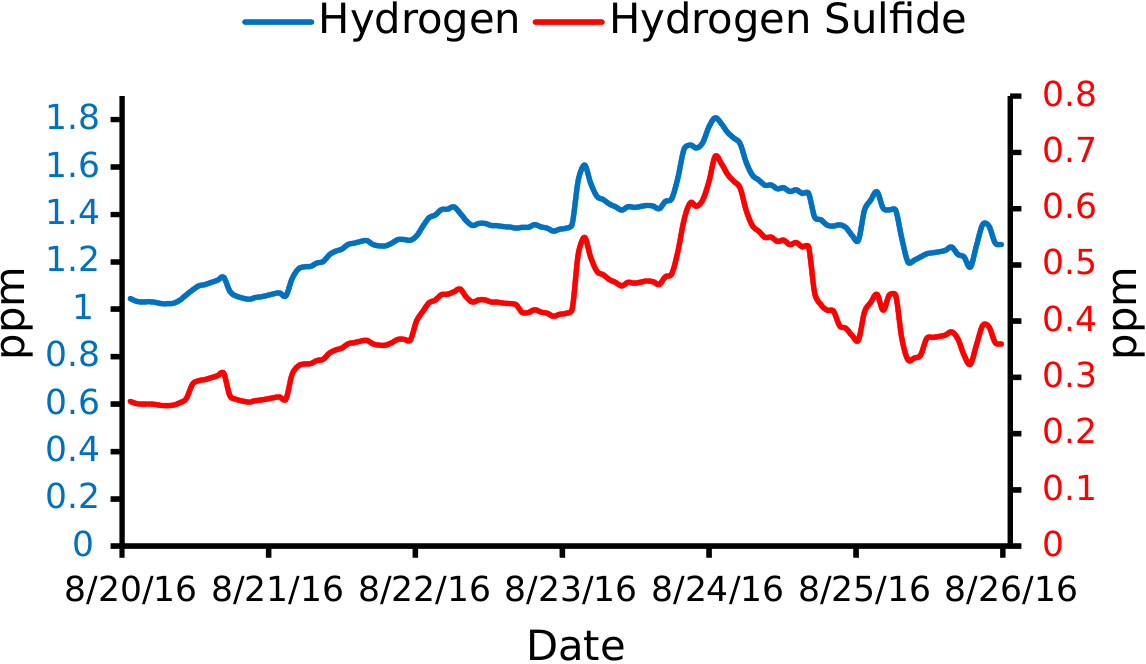}
	\caption{Hydrogen (ppm) and Hydrogen Sulfide (ppm) levels by waspmotes.}
	\label{fig_4b}
\end{figure}

\begin{figure}[h!]
	\centering
	\includegraphics[width=0.45\textwidth]{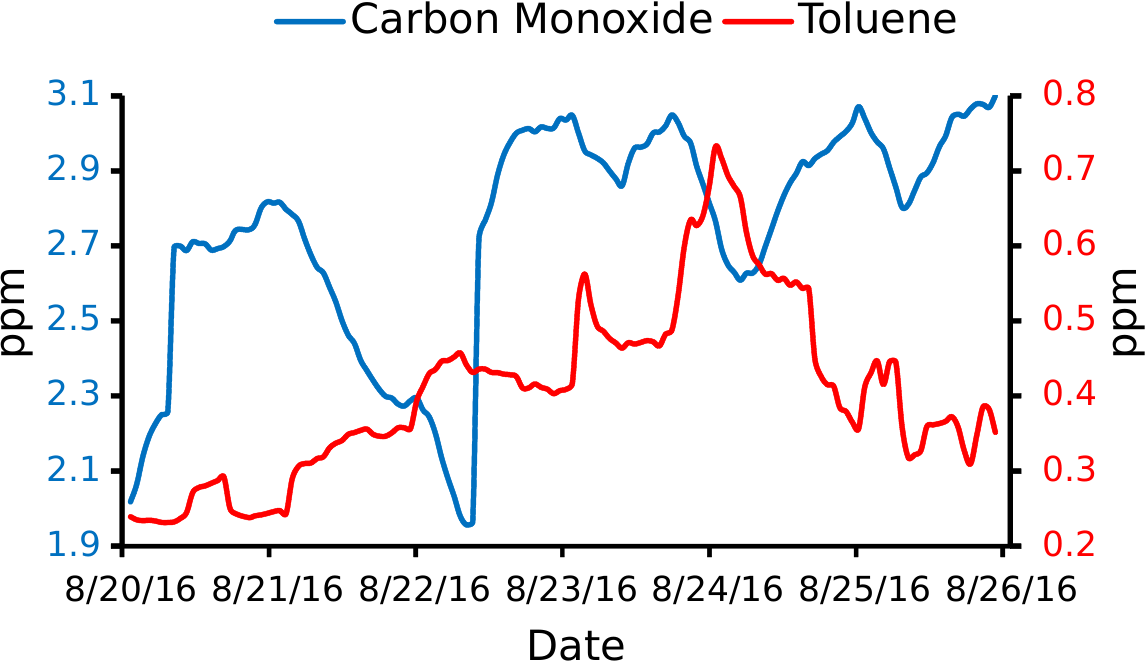}
	\caption{Carbon Monoxide (ppm) and Toluene (ppm) levels by waspmotes.}
	\label{fig_4c}
\end{figure}

The case study in this paper was intrigued by a slight incident of a fainting student  in a laboratory room, marked with an "x" on the floor plan, as shown in Fig.~\ref{fig_A}. During the period, {measurements of the sensor ESB.10.236 located in that room} were recorded as depicted in Fig.~\ref{fig_5} for temperature and humidity,  as well as in Figs.~\ref{fig_4}-\ref{fig_4c} for the room air quality, where it can be seen from the logged data of a critical episode on the 24$^{th}$ of August 2016 with an initial assessment as lack of oxygen. After a thorough investigation, the cause of the incident turned out to be high levels of indoor air pollutants on that day. This has motivated us of the development of an enhanced indoor air quality index to forecast to the occupants to avoid experiencing severe adverse health effects.


\section{IAQI data fusion framework}
In this section we provide a brief description of data fusion to calculate Indoor Air Quality Index (IAQI) and the proposed Enhanced Indoor Air Quality Index (EIAQI) incorporating also humidity.
\subsection{Indoor Air Quality Index (IAQI)}
\definecolor{green}{rgb}{0.0,1.0,0.0}
\definecolor{yellow}{rgb}{1.0, 1.0, 0.0}
\definecolor{orange}{rgb}{1.0, 0.5, 0.0}
\definecolor{red}{rgb}{1.0, 0.0, 0.0}
\definecolor{purple}{rgb}{0.56, 0.24, 0.59}
\definecolor{purpledark}{rgb}{0.5, 0.0, 0.5}
\definecolor{maroon}{rgb}{0.69, 0.19, 0.38}
\definecolor{mahogany}{rgb}{0.75, 0.25, 0.0}
\definecolor{maroondark}{rgb}{0.5, 0.0, 0.0}
\begin{table*}[h!]
	\renewcommand{\arraystretch}{1.3}
	\caption{\textsc{Indoor air quality index (IAQI)}}
	\label{table_a}
	\centering
	\footnotesize
	\scalebox{1}{
		\begin{tabular}{c||c||c||c||c||c||c||c||c||c}
			\hline\hline
			\bfseries  CO & \bfseries CO$_2$     & \bfseries H$_2$  &  \bfseries NH$_3$& \bfseries C$_2$H$_6$O &  \bfseries H$_2$S  & \bfseries C$_7$H$_8$ & \bfseries O$_2$     & \bfseries IAQI  & \bfseries Health effects\\   
			(ppm) & (ppm) & (ppm) & (ppm) & (ppm) & (ppm) & (ppm) & (\%)  &  &  \\
			\hline
			0-0.2 & 0-379 & 0-1 & 0-24  & 0-0.49  & 0-0.00033 & 0-0.0247 & 20.95 &\cellcolor{green} 0-50 & \cellcolor{green} Good\\
			\hline  
			0.21-2 & 380-450  & 1.1-2 & 25-30  & 0.5-10  & 0.00034-1.5& 0.0248-0.6 & 19-20.9 &\cellcolor{yellow} 51-100 & \cellcolor{yellow} Moderate\\
			\hline
			2.1-9 & 451-1000  & 2.1-3 & 31-50  & 11-49   & 1.6-5 & 0.7-1.6 & 15-19 &\cellcolor{orange} 101-150 & \cellcolor{orange} Unhealthy for Sensitive \\
			\hline
			9.1-15.4 & 1001-5000  & 3.1-5 & 51-100  & 50-100  & 6-20 & 1.7-9.8 & 12-15 &\cellcolor{red} 151-200  &\cellcolor{red} Unhealthy\\
			\hline
			15.5-30.4 & 5001-30000  & 5.1-8 & 101-400 &101-700  & 21-50&   9.9-12.2 & 10-12  &  \cellcolor{purple} 201-300 &\cellcolor{purple} Very Unhealthy\\
			\hline
			30.5-50.4 & 30001-40000  &8.1-10 & 401-500  &701-1000   & 51-100& 12.3-100 & $<$10  & \cellcolor{maroon} 301-400 & \cellcolor{maroon} Hazardous\\
			\hline
		\end{tabular}
	}
\end{table*}  
Air quality index (AQI) has been used by environment protection agencies throughout the world. It is a scale of air pollution to indicate its levels to inform people around a region to adjust their outdoor activities in avoiding the health risk of getting polluted. The AQI is calculated on a real time basis to form a numerical scale with a colour code which is classified into several specific ranges. The information of AQI is very important especially to children, elderly people and people with pre-existing conditions such as cardiovascular and respiratory diseases. However, this index is usually applied to outdoor instead of indoor environments even though the indoors such as work places, hotels, homes, bedrooms and theater halls also have a certain impact on human health.  For outdoor air quality, the AQI is calculated from a ratio introduced by the U.S. EPA in 2006 with the corresponding colour code with six categories ranging from good to hazardous  \cite{AQI_2018}, whereby air quality standards are based on common outdoor air pollutants such as ozone, particulate matters PM$_{2.5}$ and PM$_{10}$, CO, sulfur dioxide (SO$_2$) and nitrogen dioxide (NO$_2$). 

This research extends the existing AQI for determining the indoor air quality. Based on the AQI breakpoints, which are available online \cite{AQI_2018}, the indoor air quality index  can be evaluated with a sensing system \cite{Kim_2014}. A review of standards and guidelines for the IAQ parameters are given in \cite{Abdul-Wahab_2015}. Besides the six concentrations for the AQI as mentioned above, additional pollutants are needed  \cite{Saad_2017} to calculate the indoor air quality index (IAQI). These include carbon dioxide (CO$_2$), volatile organic compounds (VOCs), radon and formaldehyde, which are known to cause concerns of health risk \cite{Wang_2008}. For example, the hydrogen sulfide breakpoint is set in accordance with the health effects of respiratory exposure \cite{Snyder_1995}. Similarly, toluene, a toxic solvent, together with other contaminants such as formaldehyde can build up in a poorly-ventilated indoor environment. Its effect at different concentrations is explained in \cite{Baelum_1990} with breakpoint details given in \cite{VOC_2002}.  Ethanol vapour may cause irritation of the nose and throat with choking and coughing, depending on the level of concentration in air \cite{national_2011}. Ammonia, of which level breakpoint is defined in \cite{OSHA_2018}, may cause more severe problems with eyes, nose, throat and respiratory tract. High concentrations of hydrogen can cause oxygen deficit, which in turns may result in giddiness, mental confusion, loss of judgment, loss of coordination, weakness, nausea, fainting, or even loss of consciousness.
Explanation of breakpoints for hydrogen concentration can be found in \cite{Huang_2010} and for oxygen level in \cite{CCOHS_2018}. In summary, Table~\ref{table_a} lists these gases together with the IAQI in association with their health effects coded in colour.

The air quality index  for outdoor or indoor air pollutants can be calculated by using the following linear interpolation formula: 
\begin{equation}\label{eq_1}
\begin{aligned}
I_p=I_{ll}+\Big((C_p-BP_{ll})\times\frac{I_{ul}-I_{ll}}{BP_{ul}-BP_{ll}}\Big),
\end{aligned}
\end{equation}
where $I_p$ is the index for pollutant $p$, 
$C_p$ is its rounded concentration, 
$BP_{ul}$ ($BP_{ll}$) is the breakpoint greater (less) than or equal to $C_p$, and
$I_{ul}$ ($I_{ll}$) is the index value corresponding to $BP_{ul}$ ($BP_{ll}$).

In the case of oxygen level, the IAQI is calculated using the following linear interpolation formula:
\begin{equation}\label{eq_2}
\begin{aligned}
I_o=I_{ul}-\Big((BP_{ul}-C_o)\times\frac{I_{ll}-I_{ul}}{BP_{ll}-BP_{ul}}\Big),
\end{aligned}
\end{equation}
where
$I_o$ is the index for oxygen,
$C_o$ is its rounded concentration in percentage, $BP_{ul}$ ($BP_{ll}$) is the breakpoint greater (less) than or equal to $C_o$, correspondingly with the upper ($I_{ul}$) and lower ($I_{ll}$) index of oxygen.

For example, the indoor waspmote gave $C_p$=230.4295 $ppm$ for CO$_2$. We then obtained from Table~\ref{table_a} as  $BP_{ul}$ = 379, $BP_{ll}$ = 0, $I_{ul}$ = 50,  $I_{ll}$ = 0, and the IAQI obtained from (1) is  30.3997, which is in the "good" category. Now if waspmote readings for O$_2$ is $C_o$=19.7347 $\%$, the breakpoints found from the table are then $BP_{ul}$=20.9, $BP_{ll}$=19, $I_{ul}$=100, and $I_{ll}$=51. The IAQI from (2) is therefore 69.9475, which is "moderate" in health concerns.

{Eight pollutant profiles are extracted from the waspmote sensor to calculate the IAQI correspondingly. To integrate also humidity and temperature for formulating the proposed enhanced indoor air quality index (EIAQI) we consider next the humidity index.}


\subsection{Humidex}
Since the evaporation process of sweat for cooling down a human body in hot weather usually stops when the relative humidity reaches about 90\%, indoor heat may yield a rise in the body temperature, causing illness. To describe the hot or cold feelings of an average person during different seasons, Canadian meteorologists proposed the humidex a dimensionless quantity based on the dew point theory, combining the effect of heat and humidity with breakdowns given in \cite{CCOHS_2018}. Accordingly, the humidex is calculated as,
\begin{equation}\label{eq_a}
\begin{aligned}
h=T+\frac{5}{9}\times\Big(6.112\times 10^{7.5\times \frac{T}{237.7+T}}\times \frac{RH}{100}-10\Big),
\end{aligned}
\end{equation}
where $T$ is air temperature in \textdegree C and $RH$ is relative humidity in \%.  Humidex ratings can be summarized in Table~\ref{table_i}.
\begin{table}[h!]
	\renewcommand{\arraystretch}{1.3}
	\caption{\textsc{ Humidex Ratings}}
	\label{table_i}
	\centering
	\footnotesize
	\begin{tabular}{c||c } 
		\hline
		\hline
		\bfseries Humidex Range & \bfseries Degree of Comfort\\
		\hline
		\cellcolor{green} 16-29 & \cellcolor{green}  Comfort  \\	
		\hline
		\cellcolor{yellow}  30-39 & \cellcolor{yellow} No Comfort   \\
		\hline
		\cellcolor{orange} 40-45&  \cellcolor{orange} Some Discomfort  \\
		\hline
		\cellcolor{red} 46-54  & \cellcolor{red}  Great Discomfort \\
		\hline
		\cellcolor{purple} 55-60 & \cellcolor{purple}  Dangerous \\
		\hline
		\cellcolor{maroon} 61-65 &  \cellcolor{maroon} Heat Stroke \\
		\hline
	\end{tabular}
\end{table} 

\subsection{Enhanced Indoor Air Quality Index (EIAQI)}
\begin{figure}[h!]
	\centering
	\includegraphics[width=0.5\textwidth]{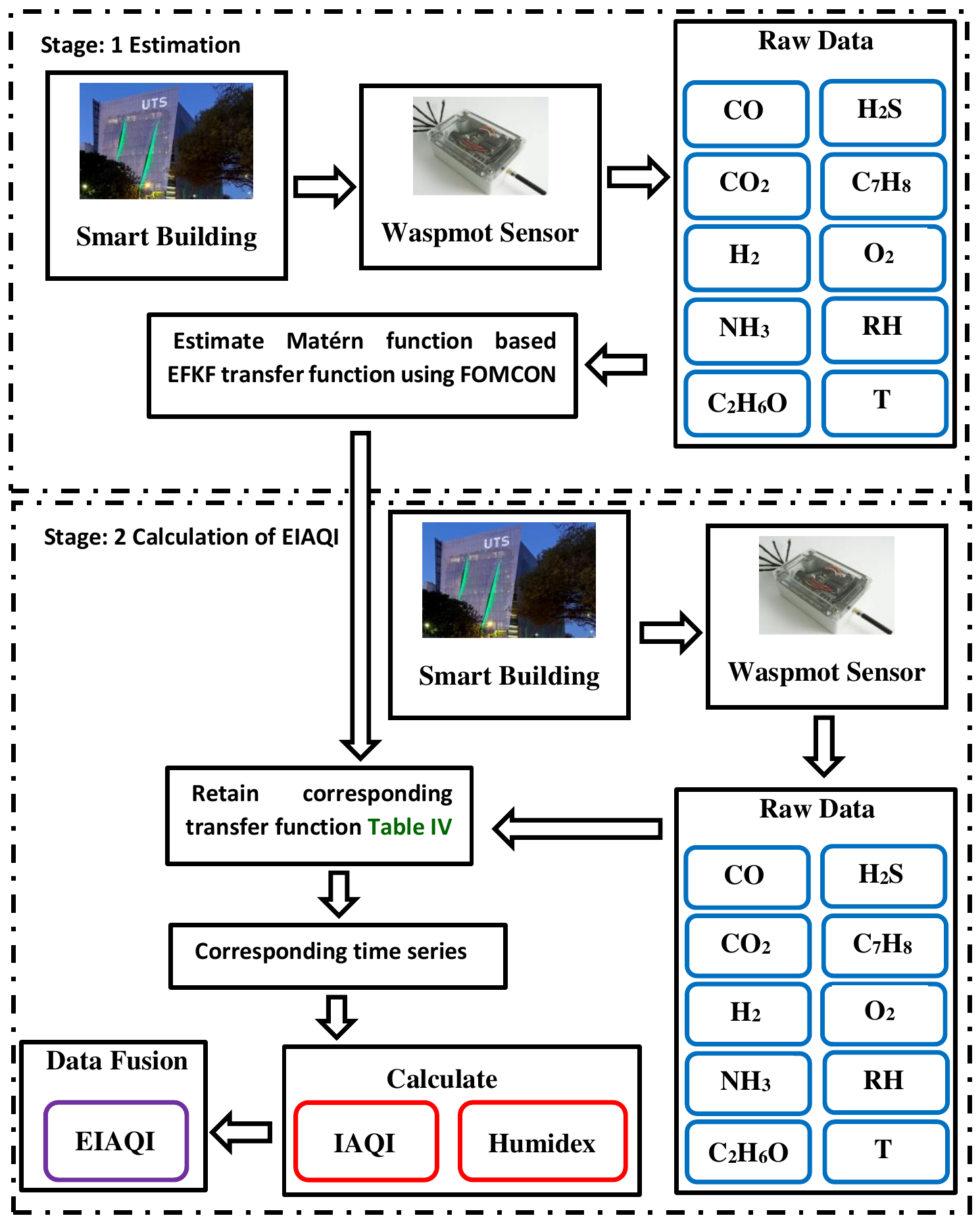}
	\caption{Flowchart of calculating EIAQI using Mat\'{e}rn covariance function based EFKF.}
	\label{fig_B}
\end{figure}
\begin{figure*}[h!]
	\centering
	\includegraphics[width=1\textwidth]{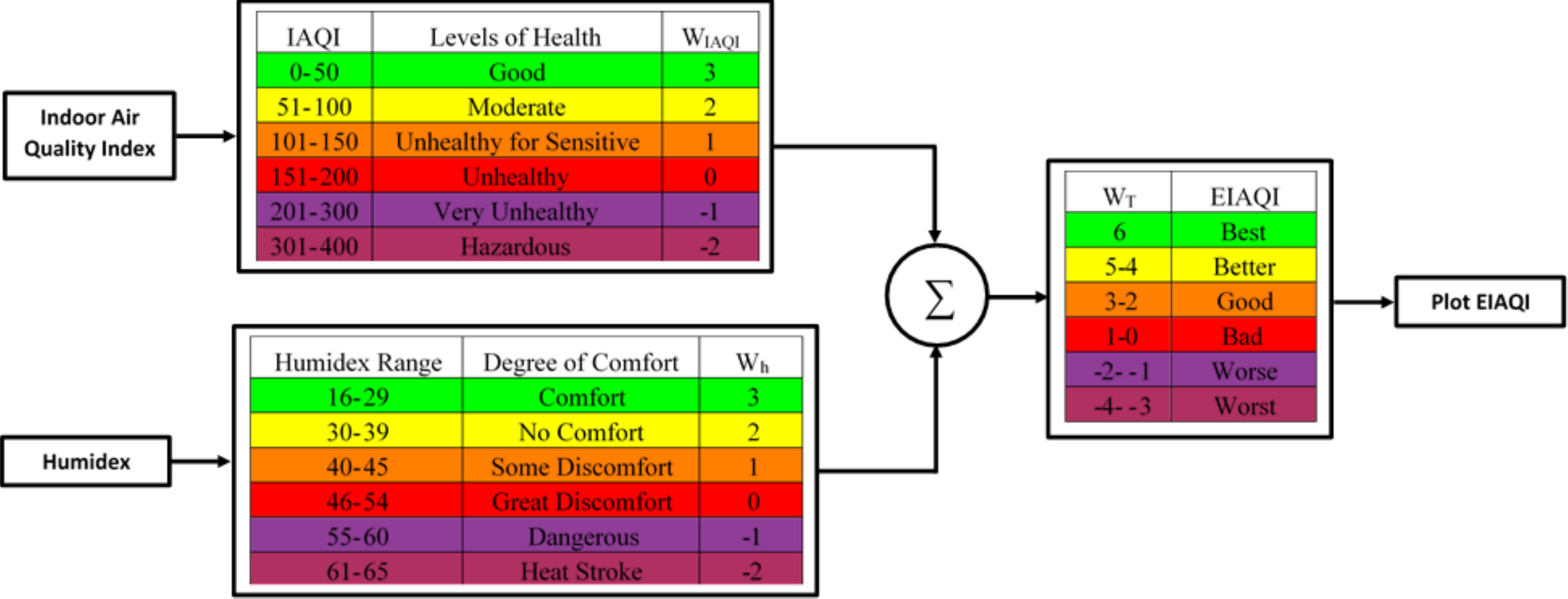}
	\caption{Calculation procedures for EIAQI.}
	\label{fig_8}
\end{figure*}
{Proposed methodology is shown as flowchart in Fig.~\ref{fig_B}. Herein, Stage 1 is based on estimation of EFKF transfer function based on Mat\'{e}rn covariance function and Stage 2 is calculate the value of EIAQI.} 

For decades, the American Society of Heating, Refrigerating and Air-Conditioning Engineers (ASHRAE) Standard 55 has been using the Fanger's predicted mean vote (PMV) model to evaluate the indoor thermal comfort satisfaction \cite{ Hu_2018}. PMV is based on the average vote of a large group of people on the a seven-point thermal sensation scale, using canonical thermal comfort models. Attempt to extend the IAQI to incorporate a thermal comfort index taking into account humidity can be found in \cite{Rana_2017}. In this work, we propose to improve the indoor air quality index by complementing it with humidex to formulate the EIAQI as,
\begin{equation}\label{eq_kk}
\begin{aligned}
EIAQI=&(W_h\times h)+(W_{IAQI}\times IAQI),\\
W_T=&W_h+W_{IAQI},
\end{aligned}
\end{equation}
where $W_h$  and $W_{IAQI}$ are respectively the humidex and  IAQI weighting factors ranging from -2 to 3, $W_T$ is the overall EIAQI weightage, $IAQI$ is the indoor air quality index, and $h$ is the humidex. The calculation procedure for the EIAQI is shown in Fig.~\ref{fig_8}. 
For example, at any given time, the status of IAQI is "Good" (weightage 3) and the Humidex status is "No Comfort' (weightage 2), then the total EIAQI weightage is 5 which refers the overall condition of the room as "Better". 

\section{Extended Fractional Kalman Filter}
A majority of research work in indoor air quality is to obtain a mathematical model based on a given set of parameters and other information of geometry, shape, size, and contrast, see e.g., \cite{Wahid_2013} to predict the pollutant distribution.
On the other hand, inverse modelling generally focuses on the mathematical process of estimating the sources when determining the spatiotemporal distribution via a set of data or observations, see e.g., \cite{Metia_2018} for an outdoor emissions problem. For indoor applications, here extended fractional Kalman filtering  is used to obtain air pollutant profiles in a smart building for accurately predicting the IAQ that the waspmotes installed in the building may overlook.

\subsection{EFKF Estimation Scheme}
The EFKF is particularly suitable for accurate and effective state estimation of highly nonlinear systems, where additive uncertainties, initial deviation, noise, disturbance and inevitably missing measurements affect the prediction performance \cite{Hu_2018a}. In outdoor air quality modelling, an EFKF with Mat\'{e}rn function-based covariances has been applied for pollutant prediction \cite{Metia_2016} to improve accuracy of inventories and to complement missing data taking into account the spatial distribution of the indoor air quality  profiles. Here, by adopting a Mat\'{e}rn correlation function for a length scale $l=\sqrt 5/\lambda$, the EFKF of fractional order $\alpha$ is proposed as
\begin{align}\label{eq_3}
\frac{d^{\alpha}f(t_k)}{dt^{\alpha}}&=\begin{bmatrix}
0 & 1 & 0 & 0\\
0 & 0 & 1 & 0 \\
0 & 0 & 0 & 1\\
-\lambda^4 & -4\lambda^3 & -6\lambda^2 & -4\lambda\\
\end{bmatrix}f(t_k)
+ \begin{bmatrix}
0 \\
0 \\
0 \\
1 \\
\end{bmatrix}w(t_k),\nonumber \\
y(t_k)&=\begin{bmatrix}
1 & 0 & 0 & 0
\end{bmatrix}f(t_k)+d(t_k),
\end{align} 
where $\lambda$ is a positive constant for the system quadruple pole (at $-\lambda$) depending on the correlation length $l$ of the Gaussian process involved \cite{Metia_2016}, $f(t_k)$ represents waspmote data assumed to have initial zero mean and covariance matrix $diag\{0.1\}$ with measurement variance $0.5^2$, spectral density of process noise $10^{-6}$ and {$k$ is discrete-time index.} 

\subsection{Fractional Order Identification}
Fractional-order systems are considered as a generalization of integer-order ones to improve system performance. In this work, our implementation is based on the Fractional-Order Modeling and Control (FOMCON) Toolbox in MATLAB \cite{ book:Tepljakov_2017} with data collected in the time domain from waspmotes. Air pollutant concentrations, after conversion, are to be processed for prediction of abnormalities by using the EFKF where the fractional order is identified with FOMCON. 
Here, the black box modelling \cite{book:Ljung_2011} is applied to infer a dynamic system model based upon experimentally collected data. This filtered model represents a
relationship between system inputs and outputs under external stimuli in order to determine and predict the system behavior. Let
$y_r$ denote the experimental pollutant profile using eqn. (\ref{eq_3}) as a plant output, and $y_m$ the identified model output. We consider the single-input and single-output (SISO) case where both $y_r$ and $y_m$ are $N \times 1$ vectors with the model output error:  
\begin{equation}\label{eq_9}
\epsilon=y_r-y_m,
\end{equation}
where estimation performance can be evaluated via the maximum absolute error: 
\begin{equation}\label{eq_12}
\epsilon_{max} = \underset{i}{\mathrm{max}}\mid\epsilon(i)\mid,
\end{equation}
or the mean squared error:
\begin{equation}\label{eq_13}
\epsilon_{MSE}=\frac{1}{N}\sum_{i=0}^{N}\epsilon_i^2=\frac{\norm{\epsilon}_2^2}{N},
\end{equation}
{where $N$ is the number of samples.}

To demonstrate the merit and advantage of using EFKF to estimate pollutant profiles in smart buildings, let's take the concentration of CO$_2$ on the 21$^{st}$ to 23$^{rd}$  August 2016 and $N=78$  from data of the considered building. {Three days, three hours before starting date and three hours after end date are considered for ideal case to identify transfer function. Starting date is 21$^{st}$ August 2016 and ending date is 23$^{rd}$ August 2016. The value of $N$ is 3$\times$24+3+3=78.} From conventional system identification, a corrected indoor air quality profile can be obtained from the corresponding rational transfer function as:
\begin{equation}\label{eq_14}
F(s)=\frac{1}{a_4 s^4 + a_3s^3 + a_2s^2 +a_1s + a_0}, \\ 
\end{equation}
where $a_4=1$, $a_3=1.058\times 10^{-1}$, $a_2=4.2\times 10^{-3}$, $a_1=7.408\times 10^{-5}$, and $a_0=4.9\times 10^{-7}$ for the pollutant level data collected at the testbed building. In fractional order modelling, the identification problem is included in estimating a set of parameters  $a_n=[a_4\quad a_3\quad a_2\quad a_1\quad a_0]$ and $\alpha_n=[\alpha_{a_4}\quad \alpha_{a_3}\quad \alpha_{a_2}\quad \alpha_{a_1}\quad \alpha_{a_0} ]$ for the transfer function of the model (\ref{eq_3}),
\begin{equation}\label{eq_15}
F^{\alpha}(s)=\frac{1}{a_{4}s^{\alpha_{a_{4}}}+a_{3}s^{\alpha_{a_{3}}}+a_{2}s^{\alpha_{a_{2}}}+a_{1}s^{\alpha_{a_{1}}}+a_{0}s^{\alpha_{a_{0}}}}.
\end{equation}
Table~\ref{table_ii} shows the values of fractional orders obtained by using the FOMCON toolbox with the initial transfer function from equation~(\ref{eq_14}) for all the indoor air pollutants, oxygen, temperature and humidity as collected by the building's waspmotes during the week from the 21$^{st}$ to 23$^{rd}$ of August 2016. Here, contaminant gases include CO$_2$, CO, H$_2$, NH$_3$, C$_2$H$_6$O, H$_2$S, and C$_7$H$_8$ with the corresponding mean squared error $\epsilon_{MSE}$ ranging between 0.3 and 0.9 for $N=78$.

\begin{table*}[t!]
	\renewcommand{\arraystretch}{1.3}
	\caption{\textsc{Fractional order system estimated by using FOMCON}}
	\label{table_ii}
	\centering
	\footnotesize
	\scalebox{1}{\begin{tabular}{l||c||r}
			\hline\hline
			\bf System input & \bf Fractional order system &  $\bf \epsilon_{MSE}$\\ 
			\hline
			& & \\
			Carbon Dioxide & $\frac{1}{10.297s^{3.0213}+10.463s^{1.3718}+81.103s^{1.2455}+74.212s^{1.2287}+1.0541s^{0.0046851}}$ & 0.8612\\
			& & \\
			\hline
			& & \\
			Carbon Monoxide & $\frac{1}{40.309s^{4.1246}-53.264s^{3.9793}+20.59s^{3.5975}+3.178s^{1.2639}+1.0523s^{0.0087201}}$ & 0.3993 \\
			& & \\
			\hline
			& & \\
			Oxygen & $\frac{1}{8.209s^{3.7867}+20.644s^{1.9433}+2.7378s^{1.5326}+0.215s^{1.5043}+1.142s^{0.026486}}$ & 0.7082\\
			& & \\
			\hline
			& & \\
			Hydrogen & $\frac{1}{11.996s^{3.2658}-2.1778s^{2.0137}+19.827s^{1.9881}+3.6513s^{1.2145}+1.6149s^{0.010517}}$ & 0.5122 \\
			& &\\
			\hline
			& & \\
			Ammonia & $\frac{1}{6.9802s^{2.2789}+10.622s^{2.2469}-1.8461s^{2.2268}-1.0216s^{1.915}+1.0213s^{0.00045903}}$ & 0.6103\\
			& &\\
			\hline
			& & \\
			Ethanol & $\frac{1}{-5.1094s^{4.2713}+321.5s^{2.6011}-311.47s^{2.5897}+5.5133s^{1.687}+1.0893s^{0.0057901}}$ & 0.6761\\
			& &\\
			\hline
			& & \\
			Hydrogen Sulfide & $\frac{1}{51.293s^{2.7929}-44.849s^{2.7567}+6.4413s^{1.6742}-0.47948s^{1.1401}+1.0917s^{0.0089142}}$ & 0.4738\\
			& &\\
			\hline
			& & \\
			Toluene & $\frac{1}{13.304s^{2.7828}-8.8256s^{2.5007}+8.3516s^{1.7996}-0.39703s^{1.2433}+1.0903s^{0.007752}}$ & 0.4262\\
			& &\\
			\hline
			& & \\
			Temperature & $\frac{1}{12.003s^{3.2165}+18.0826s^{2.6112}+0.0518s^{1.2501}+1.63903s^{1.4201}+0.0190s^{0.0239728}}$ & 0.3601\\
			& &\\
			\hline
			& & \\
			Humidity & $\frac{1}{-10.0314s^{3.5015}+1.8006s^{2.9507}+1.3676s^{1.8001}+1.51203s^{1.0052}+10.1093s^{0.80061078}}$ & 0.3007\\
			& &\\
			\hline
	\end{tabular}}
\end{table*} 

\subsection{Indoor Air Pollutant Profiles with EFKF}
{Flowchart is given in Fig.~\ref{fig_B}, where Stage 1 is related to identify the transfer function during 21$^{st}$ to 23$^{rd}$ of August 2016 and Stage 2 is related to calculate IAQI, humidex and EIAQI during 23$^{rd}$ to 25$^{th}$ of August 2016.}
	
\begin{figure}[h!]
	\centering
	\includegraphics[width=0.45\textwidth]{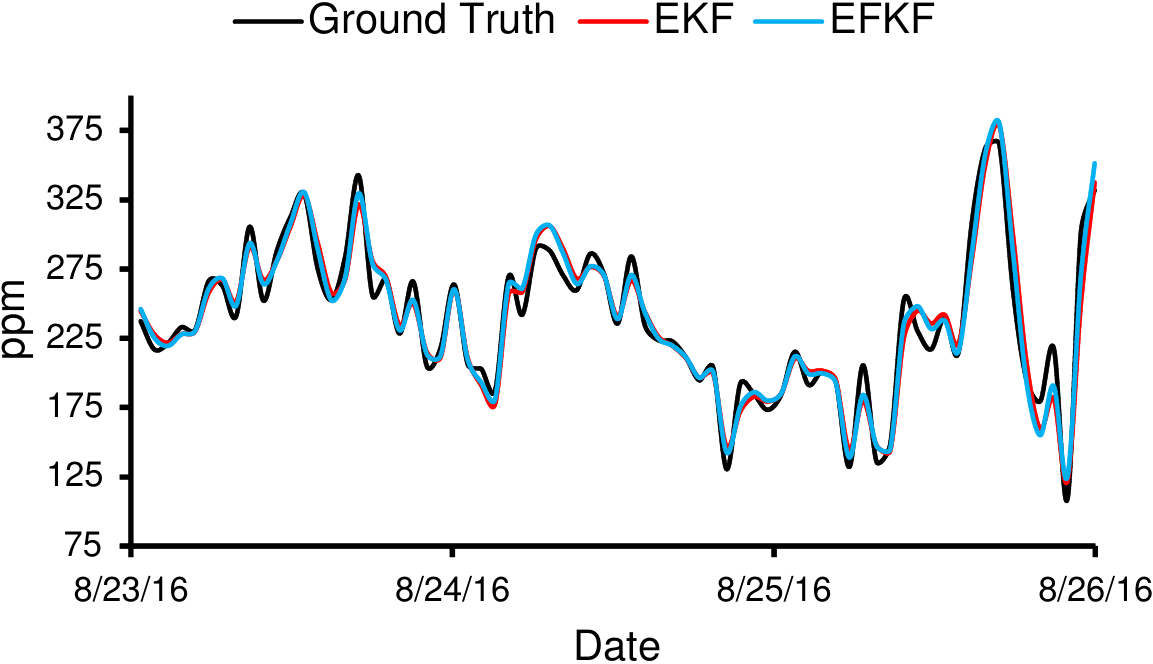}
	\caption{Carbon dioxide concentration (ppm).}
	\label{fig_10}
\end{figure}

\begin{figure}[h!]
	\centering
	\includegraphics[width=0.45\textwidth]{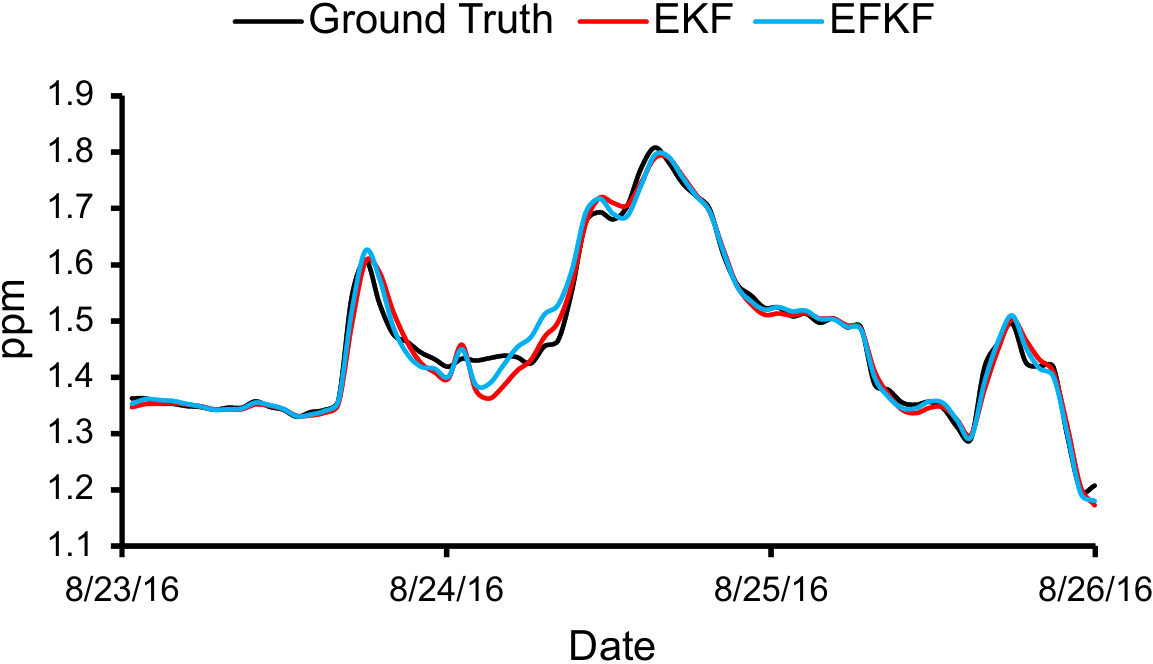}
	\caption{Hydrogen concentration (ppm).}
	\label{fig_13}
\end{figure}

\begin{figure}[h!]
	\centering
	\includegraphics[width=0.45\textwidth]{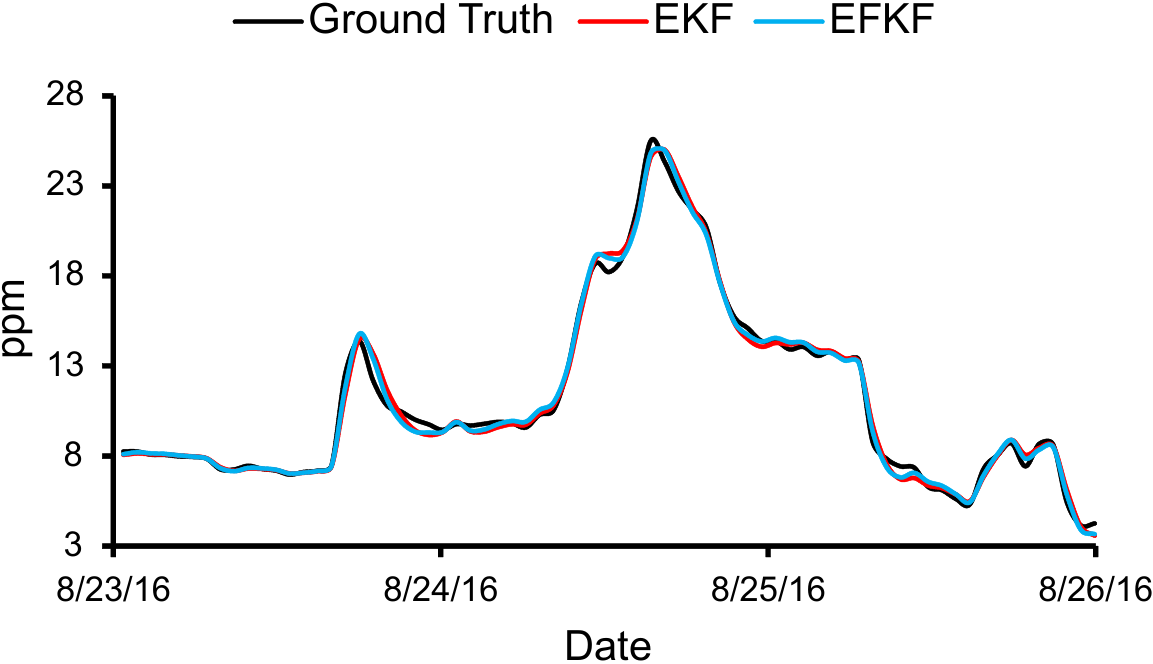}
	\caption{Ammonia concentration (ppm).}
	\label{fig_14}
\end{figure}

\begin{figure}[h!]
	\centering
	\includegraphics[width=0.45\textwidth]{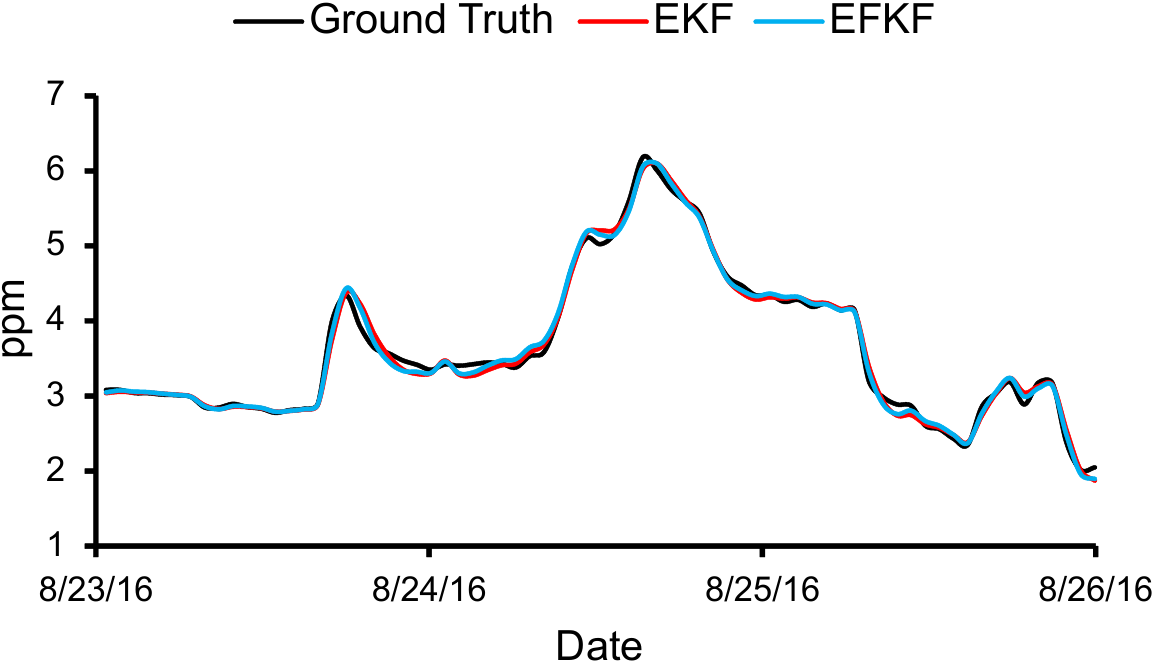}
	\caption{Ethanol concentration (ppm).}
	\label{fig_15}
\end{figure}

\begin{figure}[h!]
	\centering
	\includegraphics[width=0.45\textwidth]{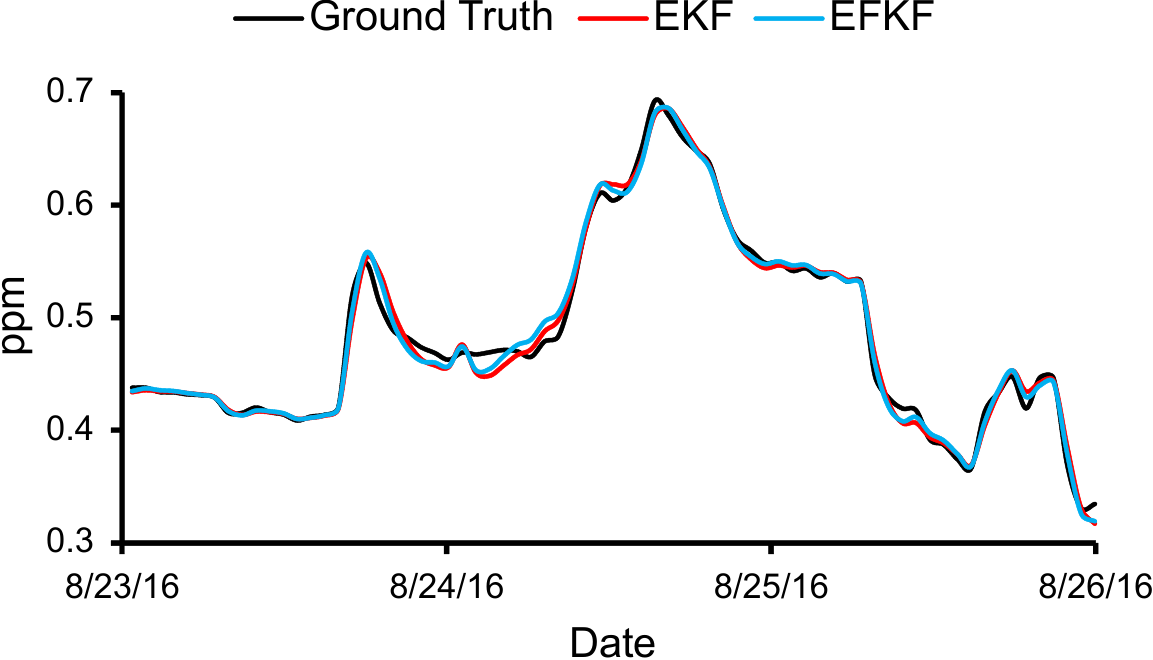}
	\caption{Hydrogen sulfide concentration (ppm).}
	\label{fig_16}
\end{figure}

\begin{figure}[h!]
	\centering
	\includegraphics[width=0.45\textwidth]{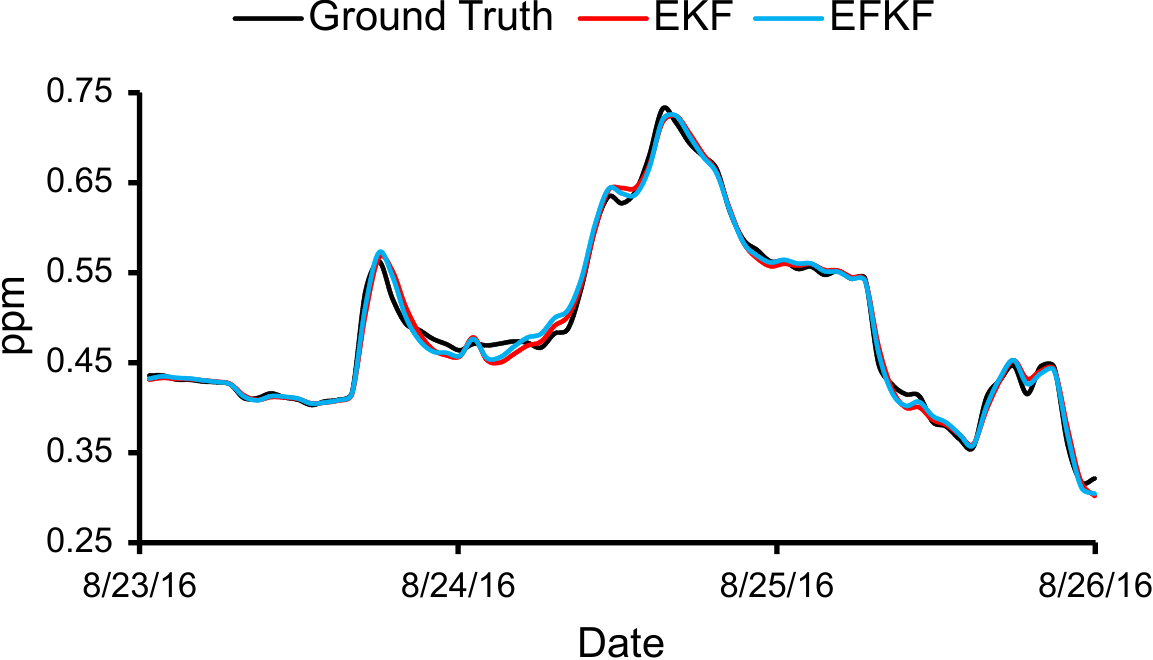}
	\caption{Toluene concentration (ppm).}
	\label{fig_17}
\end{figure}

\begin{figure}[h!]
	\centering
	\includegraphics[width=0.45\textwidth]{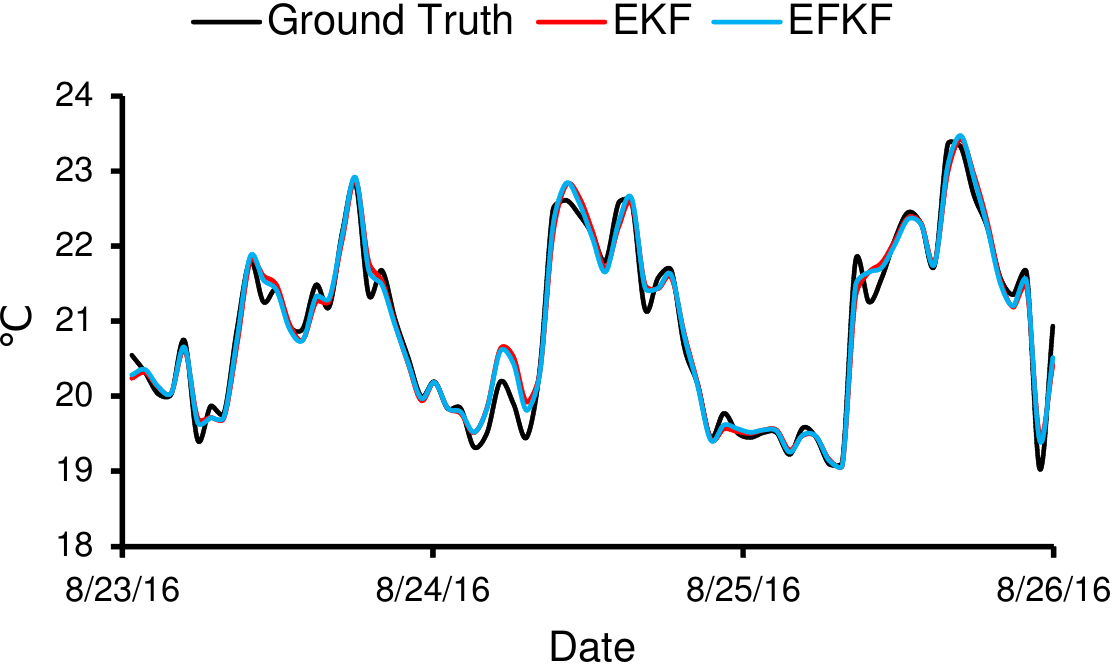}
	\caption{Temperature (\textdegree C).}
	\label{fig_18}
\end{figure}

\begin{figure}[h!]
	\centering
	\includegraphics[width=0.45\textwidth]{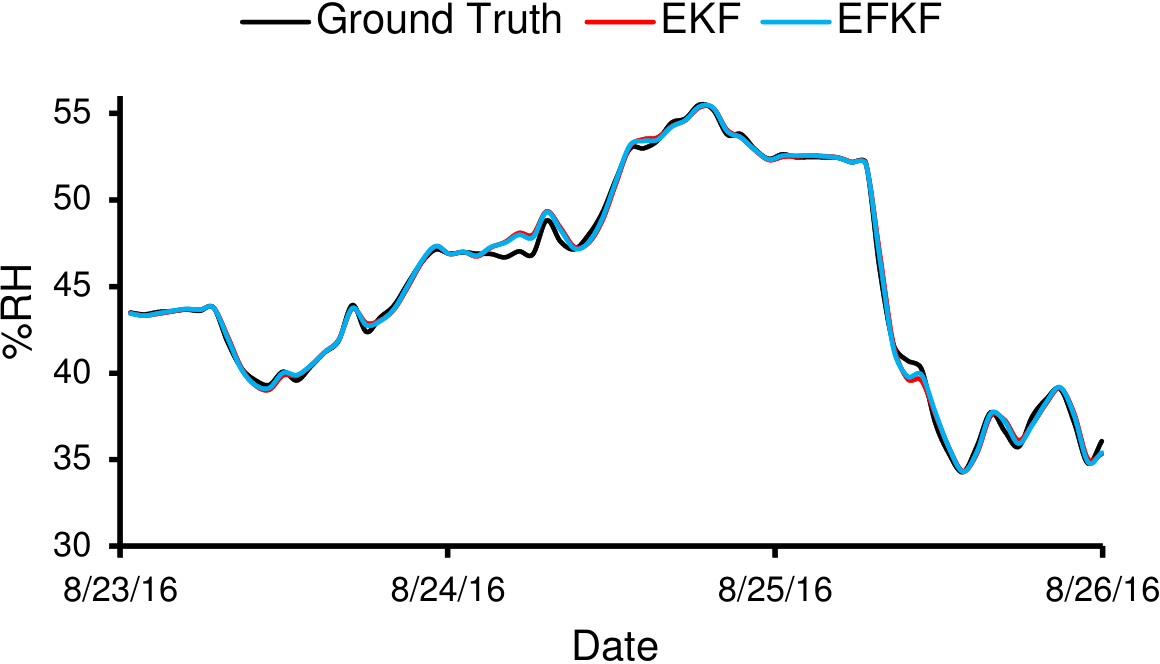}
	\caption{Humidity (\%RH).}
	\label{fig_19}
\end{figure}

\begin{figure}[h!]
	\centering
	\includegraphics[width=0.45\textwidth]{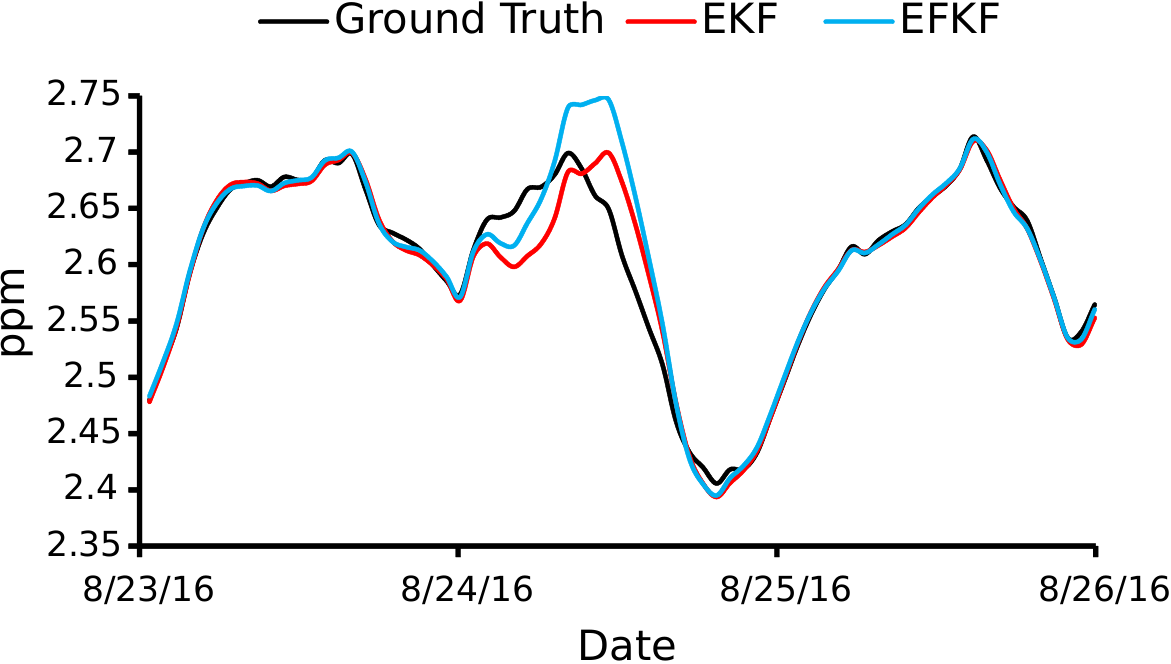}
	\caption{Carbon monoxide concentration (ppm).}
	\label{fig_11}
\end{figure}

\begin{figure}[h!]
	\centering
	\includegraphics[width=0.45\textwidth]{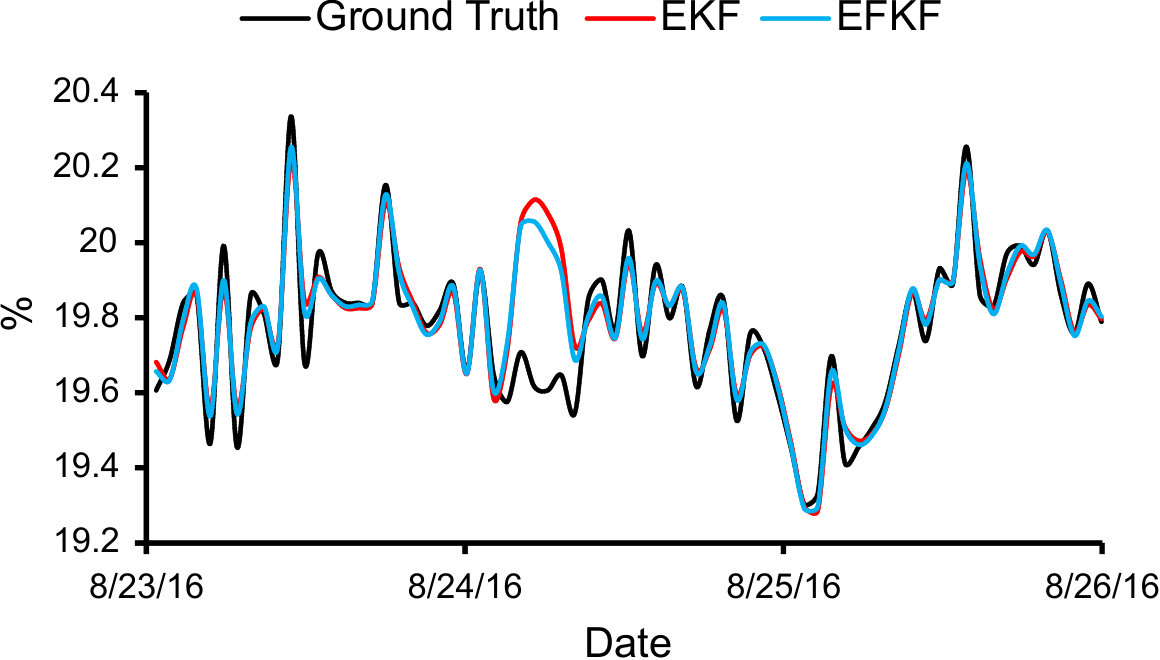}
	\caption{Oxygen concentration (\%).}
	\label{fig_12}
\end{figure}

To illustrate the improvements in determining indoor air quality profiles by using the proposed EFKF, we compare the time series of the air pollutant as well as oxygen levels over the period of interest from the 23$^{rd}$ to 26$^{th}$ of August. 
Figure~\ref{fig_10} shows the carbon dioxide concentration, distributed within a permissible limit from 400 to 1000 ppm, and rather consistent as obtained by waspmotes, EKF or EFKF. 
Similarly, the concentration distributions of gaseous contaminants such as hydrogen, ammonia, ethanol, hydrogen sulfide, toluene as well as temperature and humidity profiles are shown respectively in Figs.~\ref{fig_13}-\ref{fig_19}. They also display a general coincidence between the ground truth, EKF and EFKF. However, the carbon monoxide and oxygen levels, depicted respectively in Figs.~\ref{fig_11} and \ref{fig_12}, exhibit a difference on the 24$^{th}$ August with an increase of around 0.13 ppm in CO concentration  and 0.4\% in O$_2$ concentration by using EFKF as compared to the measured ground truth.

On one hand, while the levels of hydrogen, ammonia, ethanol and hydrogen sulfide lie in the moderate ranges as referred to Table~\ref{table_a}, the peak of these profiles all rests with the 24$^{th}$ of August, which may become unhealthy to highly sensitive people. On the other hand, the concentration of toluene C$_7$H$_8$ shows clearly a rise on the same day of over 0.7 ppm which is unhealthy for a sensitive person. Moreover, it is interesting to note that from the correction of EFKF, the level of oxygen on the incident date was moderate indeed with over 20\%, while the concentration of carbon monoxide was found rather higher than the waspmote measurements and unhealthy for sensitive people. These filtered profiles explain that the cause for the student's fainting was an exposure to not of a low oxygen concentration but of a poor indoor air quality environment with unhealthy levels of gaseous pollutants such as CO and C$_7$H$_8$, particularly in association with a substantial rise in humidity on the incident date.

\subsection{Statistical Analysis}
In order to evaluate the performance of prediction, we introduce several model performance measures including MAPE (mean absolute percentage error), RMSE (root mean square error) and R$^2$ (the coefficient of determination), defined respectively as follows:
\begin{equation}\label{eq_16}
\begin{aligned}
MAPE=\frac{100}{N}\sum_{i=1}^N\Bigg(\frac{|y_{r_i}-y_{m_i}|}{|y_{r_i}|}\Bigg),
\end{aligned}
\end{equation}
\begin{equation}\label{eq_17}
\begin{aligned}
RMSE=\sqrt{\frac{1}{N}\sum_{i=1}^N(y_{r_i}-y_{m_i})^2},
\end{aligned}
\end{equation}
\begin{equation}\label{eq_18}
\begin{aligned}
R^2=1-\Bigg(\frac{\sum_{i=1}^{N}(y_{m_i}-y_{r_i})^2}{\sum_{i=1}^{N}(y_{m_i})^2}\Bigg),
\end{aligned}
\end{equation}
where $y_{r_i}$ and $y_{m_i}$ are the forecast and observed values, and $N$ is the number of samples. MAPE and RMSE are applied as performance criteria of the prediction model to quantify the errors of forecasting values. The coefficient of determination $R^2$ is used to assess the strength of the relationship of the estimation to the accurate observation.
\begin{table*}[h!]
	\renewcommand{\arraystretch}{1.3}
	\caption{\textsc{Performance statistics of EKF and EFKF in datasets of different days}}
	\label{table_iv}
	\centering
	\footnotesize
	\centering
	\scalebox{1}{\begin{tabular}{l||c||c||c||c||c||c||c||c||c||c||c||c}
			\hline\hline
			\bf Pollutant &  \multicolumn{6}{c||}{ \bf 8/23/2016} & \multicolumn{6}{c}{ \bf 8/24/2016}\\
			\hline
			& \multicolumn{3}{c||}{EKF} & \multicolumn{3}{c||}{EFKF} & \multicolumn{3}{c||}{EKF} & \multicolumn{3}{c}{EFKF}\\
			\cline{2-13} 
			& \multicolumn{1}{c||}{MAPE} & \multicolumn{1}{c||}{RMSE} & \multicolumn{1}{c||}{R$^2$} & \multicolumn{1}{c||}{MAPE} & \multicolumn{1}{c||}{RMSE} & \multicolumn{1}{c||}{R$^2$} & \multicolumn{1}{c||}{MAPE} & \multicolumn{1}{c||}{RMSE} & \multicolumn{1}{c||}{R$^2$} & \multicolumn{1}{c||}{MAPE} & \multicolumn{1}{c||}{RMSE} & \multicolumn{1}{c}{R$^2$}\\
			\hline
			Carbon Dioxide & 2.94\% & 1.0498 &  0.9187 & 2.45\% &  0.8729 & 0.9437 & 3.31\% &  2.5957 & 0.8476 & 2.69\%& 2.2503 & 0.8982\\
			Carbon Monoxide & 0.71\% & 0.0076 & 0.9177 & 0.70\% & 0.0073 & 0.9213 & 0.94\% & 0.0078 & 0.7998 & 0.88\% & 0.0078 &  0.8202\\
			Oxygen & 0.70\% &   0.0818 & 0.8285 & 0.59\% & 0.0692 & 0.8778 & 0.52\% & 0.0631 &  0.8978 & 0.45\% & 0.0551 & 0.9229 \\
			Hydrogen & 3.40\% & 1.7091 & 0.9631 & 2.42\% & 1.3752 & 0.9761 & 3.77\% & 1.8922 &  0.9892 &  2.56\% &  1.4871 & 0.9928\\
			Ammonia & 2.20\% & 0.2681 & 0.9918 & 1.94\% &   0.2166 &  0.9946 & 3.90\% &  0.4113 & 0.9948 & 3.28\% & 0.3394 &  0.9964 \\
			Ethanol & 1.48\% & 0.0597 & 0.9942 & 1.34\% &  0.0491 &  0.9960 & 2.41\%  &  0.0911 & 0.9937 & 2.09\% & 0.0779 &  0.9954  \\
			Hydrogen Sulfide & 1.08\% &  0.0059 & 0.9946 & 1.01\% &  0.0051 & 0.9959 & 1.68\% & 0.0092 & 0.9918 & 1.51\% &  0.0082 &  0.9934\\
			Toluene & 1.18\% & 0.0065 &  0.9946 & 1.10\% & 0.0055 & 0.9961 & 1.86\% & 0.0100 &  0.9925 & 1.65\% & 0.0089 & 0.9941 \\
			Temperature & 0.65\% &  0.1840 &  0.9613 & 0.56\% &  0.1579 & 0.9717 & 0.82\% & 0.2349 &  0.9852 & 0.71\% &  0.1983 & 0.9896 \\
			Humidity &0.54\% &0.2593 &  0.9987 & 0.46\% &  0.2176 & 0.9992 & 0.8\% & 0.4416 &  0.9984 & 0.67\% & 0.3757 & 0.9988 \\
			\hline
	\end{tabular}}
\end{table*} 
Table~\ref{table_iv} provides descriptive statistics of EKF and EFKF prediction data on the 23rd and 24th of August, 2016. It justifies for the improvement of EFKF over EKF in precise estimation of indoor air quality. This is accounted by the merit of the  Mat\'{e}rn covariance function associated with the Kalman filters used to allow for better correlation at a suitable length scale between the waspmotes and a location inside the building, here $\lambda=\frac{\sqrt(5)}{l}$ and $l=5m$, the distance from the corridor (waspmote) to the lab room (incident location, in this study). As can be seen, the RMSE value is higher in the case of CO$_2$ concentration as compared to other IAQ levels. This is explained by a biological factor whereby human beings also produce CO$_2$ due to the natural process of respiration with a wide range of permissible limits (400-1000 ppm).

\section{Indoor Air Quality Assessment}
The above findings indicate the importance of accurate, comprehensive and continuous monitoring with a prediction system for the IAQ, taking into account also human comfort. Such a system should be integrated into a building management for better monitoring the IAQ and, more importantly, prevention of any incidents via, e.g., ventilation control. 

\subsection{Time Series Plot of IAQI and Humidex Using Real Data and Estimated Data} 
\begin{figure}[h!]
\centering
\includegraphics[width=0.45\textwidth]{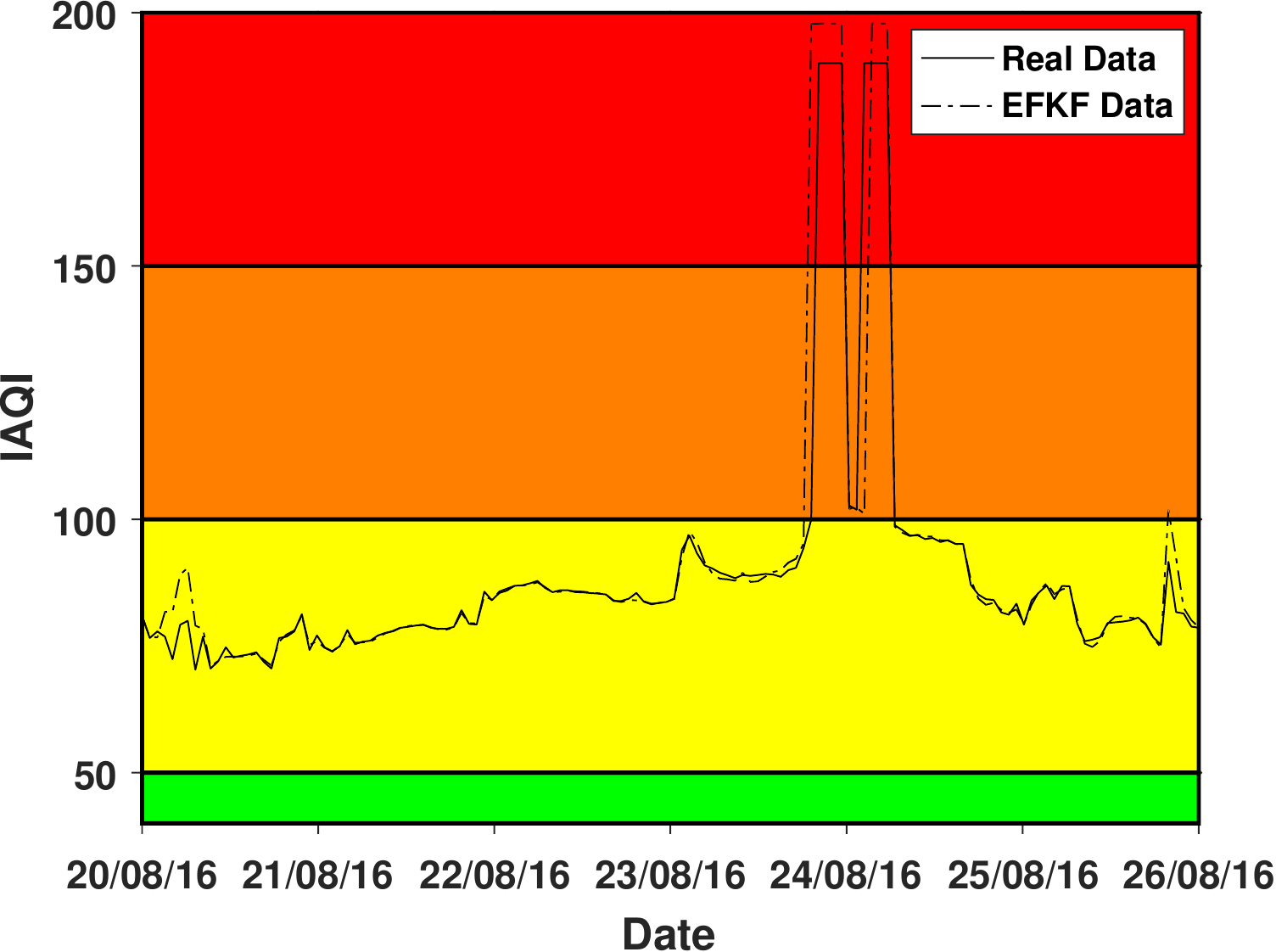}
\caption{ IAQI plot using real data and EFKF data.}
\label{fig_21}
\end{figure}

To consider the overall indoor air quality index for calculation of the IAQI, Eqn.~(\ref{eq_1}) is used to interpolate data of all pollutants except the oxygen level which is obtained from Eqn.~(\ref{eq_2}). Figure~\ref{fig_21} shows the time series plot of IAQI using real data from waspmotes and processed data from EFKF. It can be seen that the value of IAQI is rather high and particularly very unhealthy on the 24$^{th}$ August, 2016, which may affect a sensitive person. 

\begin{figure}[h!]
	\centering
	\includegraphics[width=0.45\textwidth]{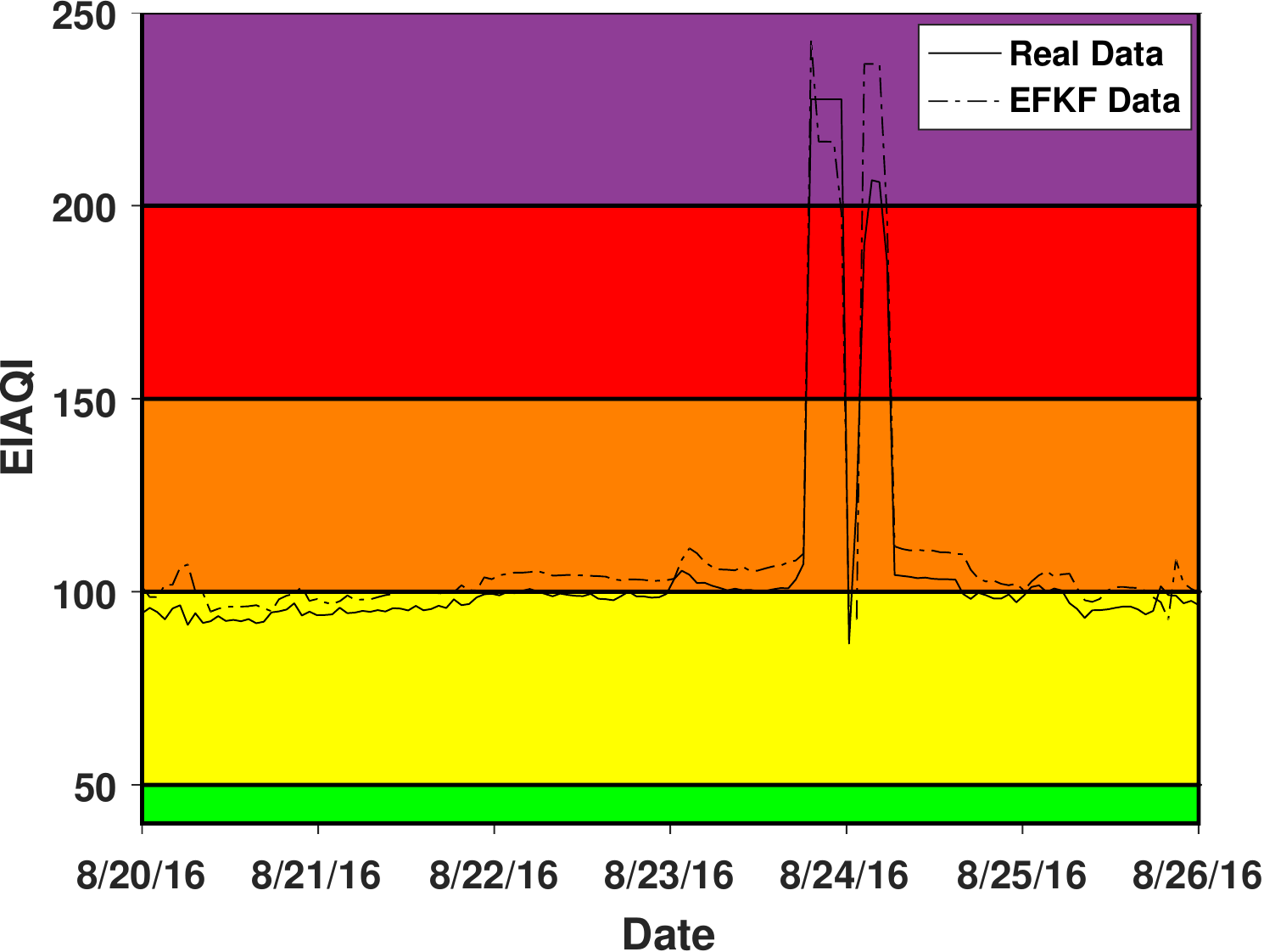}
	\caption{Overall index for indoor air using real data and EFKF data .}
	\label{fig_22}
\end{figure}
\subsection{Time Series Plot of EIAQI Using Real Data and Estimated Data} 
To incorporate also humidity, the enhanced indoor air quality index is calculated by using Eqn. ~\ref{eq_kk}. For a better illustration of the improvement obtained from the use of EFKF, the EIAQI plot is shown in Fig.~\ref{fig_22} for both real data and estimated data according to weightage $(W_h=1.0$ and $W_{IAQI}=1.0$). Using the same colour codes for health effects presented in Fig.~\ref{fig_21}, it can be seen that with EFKF, the obtained EIAQI clearly indicates an increase in the indoor air quality index within a short period of time. This is reflected in the soon recovery of the sensitive student whereas the majority of the class could tolerate. Although the proposed EIAQI with EFKF estimation is not much different with real data from waspmotes for most of the time, the enhanced indoor air quality index appears to be more accurate in reflecting the indoor air quality with EFKF data during the episode day, owing to the advantages in handling missing data as well as nonlinear and uncertain spatio-temporal distributions.

\begin{figure}[h!]
	\centering
	\includegraphics[width=0.45\textwidth]{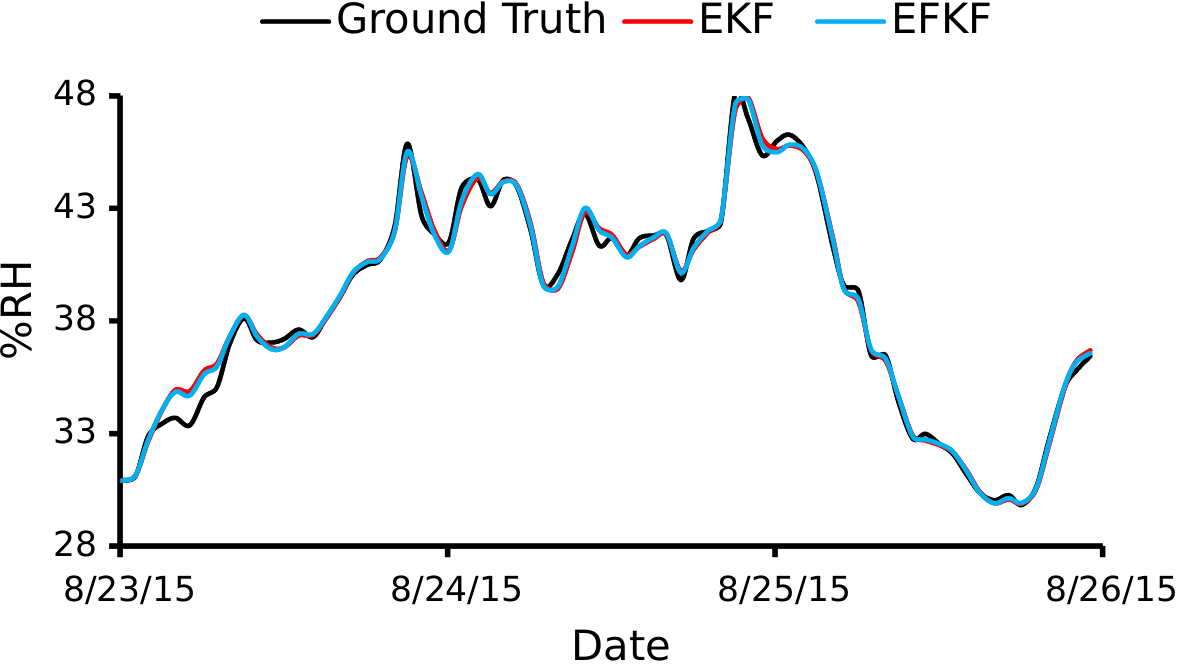}
	\caption{Humidity (\%RH).}
	\label{fig_23}
\end{figure}

\begin{figure}[h!]
    	\centering
    	\includegraphics[width=0.45\textwidth]{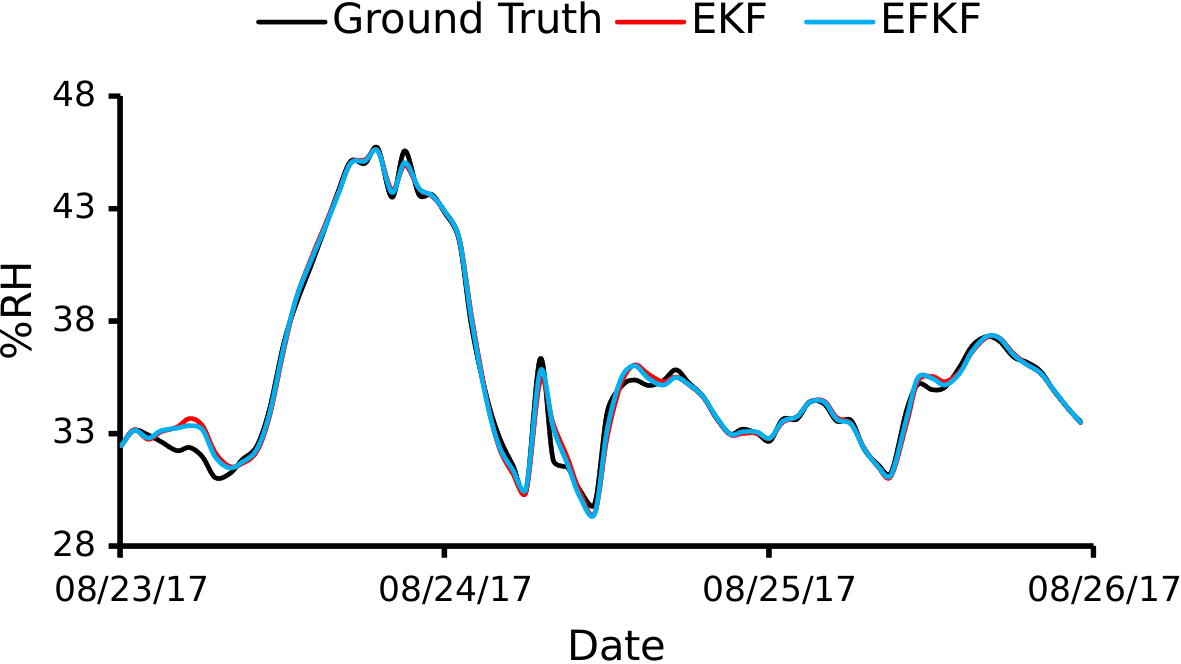}
    	\caption{Humidity (\%RH).}
    	\label{fig_24}
    \end{figure}

\begin{figure}[h!]
	\centering
	\includegraphics[width=0.45\textwidth]{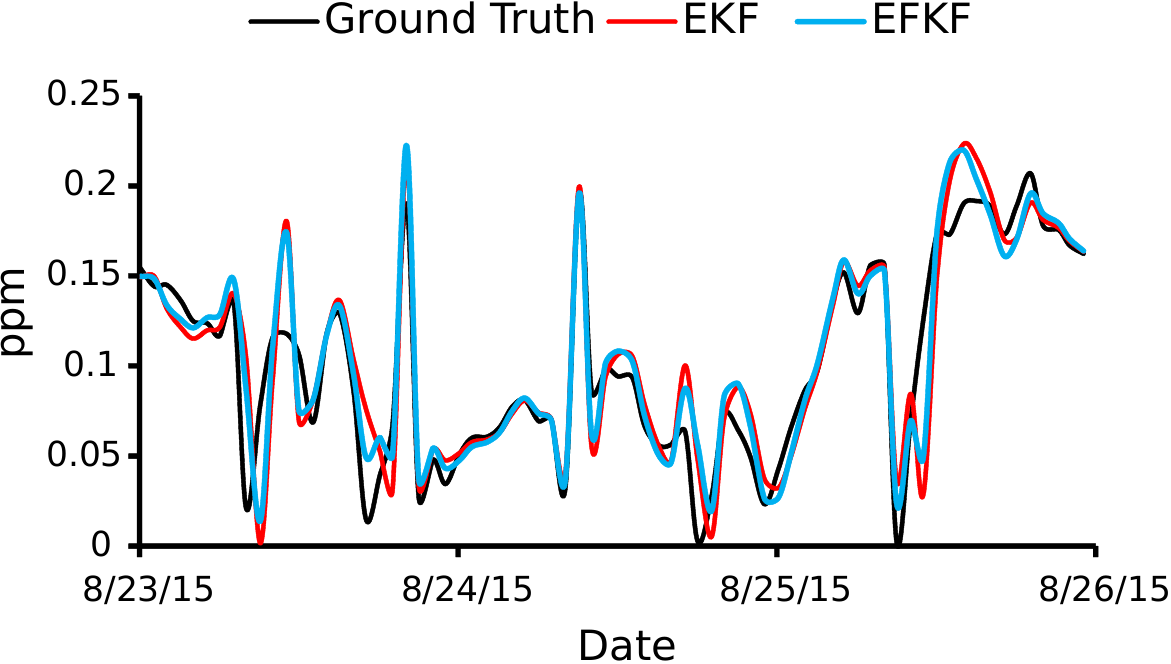}
	\caption{Carbon monoxide concentration (ppm).}
	\label{fig_25}
\end{figure}

\begin{figure}[h!]
	\centering
	\includegraphics[width=0.45\textwidth]{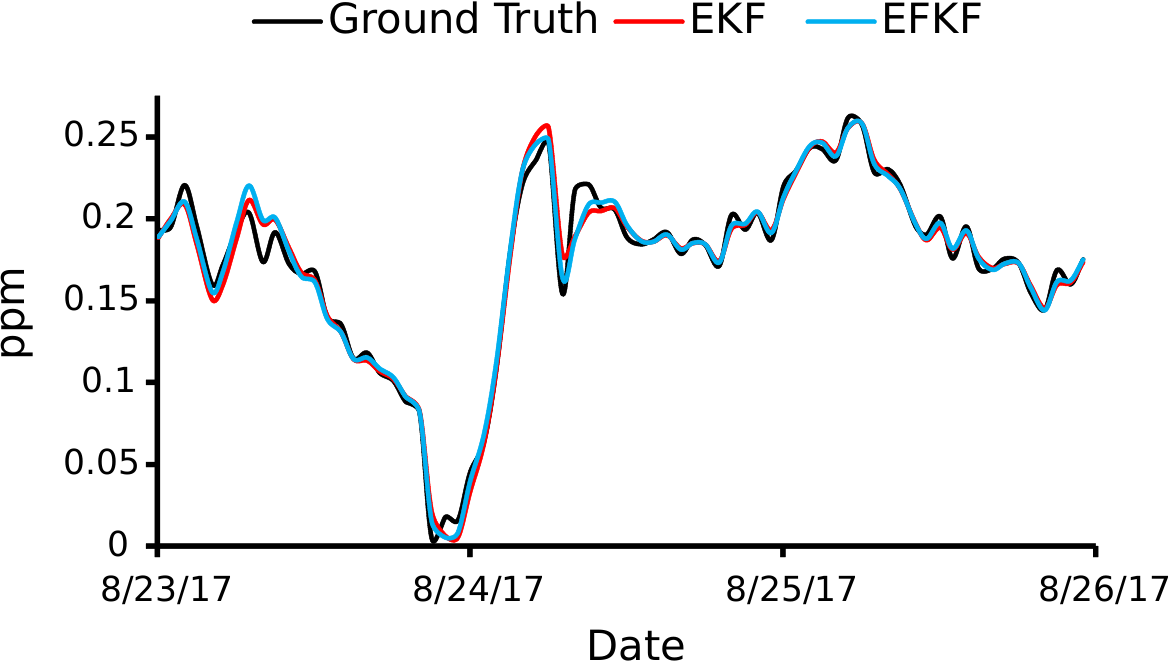}
	\caption{Carbon monoxide concentration (ppm).}
	\label{fig_26}
\end{figure}

For further comparison, we consider the same week between the 23$^{rd}$ to 25$^{th}$ of August in the previous year 2015 and the following year 2017 for humidity and carbon monoxide. Figs.~\ref{fig_23}, \ref{fig_24} and Figs.~\ref{fig_25}, \ref{fig_26} show time series plots respectively of \%RH and CO (ppm) for 2015 and 2017. It can be seen that the percentage of relative humdity and concentration of carbon monoxide in the years 2015 as well as 2017 are both smaller that those in 2016 in the same time interval. This study thus emphasises the need of a reliable management system for monitoring of indoor air pollutants beyond sensor measurements especially in a critical period, for which sensing data fusion remains indispensable to accurately assess the IAQ.

\section{Conclusion}
In this paper, we have proposed an effective approach to improve accuracy in predicting indoor air pollutant profiles taking into account their nonlinear and stochastic nature, along with a novel index for indoor air quality considering also humidity. Here, an extended Kalman filter with a fractional order is developed for the indoor air quality model, in dealing with high nonlinearity and missing or inaccurate data collected from the building's sensors. To verify the performance imrovement, both EKF and EFKF algorithms have been implemented and compared. For illustration, an incident of a student with some slight fainting, is used as a case study to not only evaluate the effectiveness of the proposed estimation framework but also to emphasize the need of integrating  accurate IAQ monitoring and prediction into the overall building management system to better maintain the inhabitants' wellbeing. In addition, a combination of IAQI and humidex is proposed to address the effect of humidity on indoor air quality.
\ifCLASSOPTIONcaptionsoff
  \newpage
\fi

\balance
\bibliography{IEEEabrv,bibi}

\begin{thebibliography}{10}
\providecommand{\url}[1]{#1}
\csname url@samestyle\endcsname
\providecommand{\newblock}{\relax}
\providecommand{\bibinfo}[2]{#2}
\providecommand{\BIBentrySTDinterwordspacing}{\spaceskip=0pt\relax}
\providecommand{\BIBentryALTinterwordstretchfactor}{4}
\providecommand{\BIBentryALTinterwordspacing}{\spaceskip=\fontdimen2\font plus
\BIBentryALTinterwordstretchfactor\fontdimen3\font minus
  \fontdimen4\font\relax}
\providecommand{\BIBforeignlanguage}[2]{{%
\expandafter\ifx\csname l@#1\endcsname\relax
\typeout{** WARNING: IEEEtran.bst: No hyphenation pattern has been}%
\typeout{** loaded for the language `#1'. Using the pattern for}%
\typeout{** the default language instead.}%
\else
\language=\csname l@#1\endcsname
\fi
#2}}
\providecommand{\BIBdecl}{\relax}
\BIBdecl

\bibitem{Jaimini_2017}
U.~Jaimini, T.~Banerjee, W.~Romine, K.~Thirunarayan, A.~Sheth, and M.~Kalra,
  ``Investigation of an indoor air quality sensor for asthma management in
  children,'' \emph{IEEE Sensors Letters}, vol.~1, no.~2, pp. 1--4, April 2017.

\bibitem{Lay-Ekuakille_2011}
A.~{Lay-Ekuakille} and A.~{Trotta}, ``Predicting voc concentration
  measurements: Cognitive approach for sensor networks,'' \emph{IEEE Sensors
  Journal}, vol.~11, no.~11, pp. 3023--3030, Nov 2011.

\bibitem{Tang_2012}
T.~Tang, J.~Hurra{\ss}, R.~Gminski, and V.~Mersch-Sundermann, ``Fine and
  ultrafine particles emitted from laser printers as indoor air contaminants in
  german offices,'' \emph{Environmental Science and Pollution Research},
  vol.~19, no.~9, pp. 3840--3849, Nov 2012.

\bibitem{Sahin_2016}
O.~A. Sahin, N.~Kececioglu, M.~Serdar, and A.~Ozpinar, ``The association of
  residential mold exposure and adenotonsillar hypertrophy in children living
  in damp environments,'' \emph{International Journal of Pediatric
  Otorhinolaryngology}, vol.~88, no. Supplement C, pp. 233 -- 238, 2016.

\bibitem{Jin_2018}
M.~{Jin}, N.~{Bekiaris-Liberis}, K.~{Weekly}, C.~J. {Spanos}, and A.~M.
  {Bayen}, ``Occupancy detection via environmental sensing,'' \emph{IEEE
  Transactions on Automation Science and Engineering}, vol.~15, no.~2, pp.
  443--455, April 2018.

\bibitem{Zimmermann_2018}
L.~Zimmermann, R.~Weigel, and G.~Fischer, ``Fusion of nonintrusive
  environmental sensors for occupancy detection in smart homes,'' \emph{IEEE
  Internet of Things Journal}, vol.~5, no.~4, pp. 2343--2352, Aug 2018.

\bibitem{Kim_2012}
M.~Kim, B.~SankaraRao, O.~Kang, J.~Kim, and C.~Yoo, ``{Monitoring and
  prediction of indoor air quality (IAQ) in subway or metro systems using
  season dependent models},'' \emph{Energy and Buildings}, vol.~46, pp. 48 --
  55, 2012.

\bibitem{Vakiloroaya_2013}
Q.~Ha and V.~Vakiloroaya, ``Modeling and optimal control of an energy-efficient
  hybrid solar air conditioning system,'' \emph{Automation in Construction},
  vol.~49, pp. 262--270, 2015.

\bibitem{Sun_2013}
B.~{Sun}, P.~B. {Luh}, Q.~{Jia}, Z.~{Jiang}, F.~{Wang}, and C.~{Song},
  ``Building energy management: Integrated control of active and passive
  heating, cooling, lighting, shading, and ventilation systems,'' \emph{IEEE
  Transactions on Automation Science and Engineering}, vol.~10, no.~3, pp.
  588--602, July 2013.

\bibitem{Zhao_2018}
L.~Zhao, J.~Liu, and J.~Ren, ``{Impact of various ventilation modes on IAQ and
  energy consumption in Chinese dwellings: First long-term monitoring study in
  Tianjin, China},'' \emph{Building and Environment}, vol. 143, pp. 99 -- 106,
  2018.

\bibitem{Merabtine_2018}
A.~Merabtine, C.~Maalouf, A.~A.~W. Hawila, N.~Martaj, and G.~Polidori,
  ``{Building energy audit, thermal comfort, and IAQ assessment of a school
  building: A case study},'' \emph{Building and Environment}, vol. 145, pp. 62
  -- 76, 2018.

\bibitem{Rana_2017}
R.~Rana, B.~Kusy, R.~Jurdak, J.~Wall, and W.~Hu, ``Feasibility analysis of
  using humidex as an indoor thermal comfort predictor,'' \emph{Energy and
  Buildings}, vol.~64, pp. 17 -- 25, 2013.

\bibitem{Wolkoff_2018}
P.~Wolkoff, ``Indoor air humidity, air quality, and health - an overview,''
  \emph{International Journal of Hygiene and Environmental Health}, vol.~3, pp.
  376--390, Apr 2018.

\bibitem{Assa_2015}
A.~Assa and F.~Janabi-Sharifi, ``A kalman filter-based framework for enhanced
  sensor fusion,'' \emph{IEEE Sensors Journal}, vol.~15, no.~6, pp. 3281--3292,
  June 2015.

\bibitem{Carminati_2012}
M.~{Carminati}, G.~{Ferrari}, R.~{Grassetti}, and M.~{Sampietro}, ``Real-time
  data fusion and mems sensors fault detection in an aircraft emergency
  attitude unit based on kalman filtering,'' \emph{IEEE Sensors Journal},
  vol.~12, no.~10, pp. 2984--2992, Oct 2012.

\bibitem{Li_2016}
X.~Li and J.~Wen, ``System identification and data fusion for on-line adaptive
  energy forecasting in virtual and real commercial buildings,'' \emph{Energy
  and Buildings}, vol. 129, pp. 227 -- 237, 2016.

\bibitem{Wang_2017}
R.~Wang, Y.~Li, H.~Sun, and Z.~Chen, ``Analyses of integrated aircraft cabin
  contaminant monitoring network based on kalman consensus filter,'' \emph{ISA
  Transactions}, vol.~71, pp. 112 -- 120, 2017, special issue on Distributed
  Coordination Control for Multi-Agent Systems in Engineering Applications.

\bibitem{Jun_2014}
M.~Jun, ``Mat\'{e}rn-based nonstationary cross-covariance models for global
  processes,'' \emph{Journal of Multivariate Analysis}, vol. 128, pp. 134 --
  146, 2014.

\bibitem{Yue_2017}
X.~{Yue}, M.~{Kauer}, M.~{Bellanger}, O.~{Beard}, M.~{Brownlow}, D.~{Gibson},
  C.~{Clark}, C.~{MacGregor}, and S.~{Song}, ``Development of an indoor
  photovoltaic energy harvesting module for autonomous sensors in building air
  quality applications,'' \emph{IEEE Internet of Things Journal}, vol.~4,
  no.~6, pp. 2092--2103, Dec 2017.

\bibitem{Zhao_2019}
L.~{Zhao}, W.~{Wu}, and S.~{Li}, ``Design and implementation of an iot-based
  indoor air quality detector with multiple communication interfaces,''
  \emph{IEEE Internet of Things Journal}, vol.~6, no.~6, pp. 9621--9632, Dec
  2019.

\bibitem{Chang_2019}
C.~Y. {Chang}, S.~{Guo}, S.~{Hung}, and Y.~{Lin}, ``Performance analysis of
  indoor smart environmental control factors: Using temperature to control the
  rate of formaldehyde emission,'' \emph{IEEE Access}, vol.~7, pp.
  163\,749--163\,756, 2019.

\bibitem{Liu_2019}
Y.~{Liu}, K.~{Akram Hassan}, M.~{Karlsson}, O.~{Weister}, and S.~{Gong},
  ``Active plant wall for green indoor climate based on cloud and internet of
  things,'' \emph{IEEE Access}, vol.~6, pp. 33\,631--33\,644, 2018.

\bibitem{AQI_2018}
\BIBentryALTinterwordspacing
({2018}) {AQI Breakpoints}. \url{https://www.epa.gov/aqs}. [Online]. Available:
  \url{https://www.epa.gov/aqs}
\BIBentrySTDinterwordspacing

\bibitem{Kim_2014}
J.~Y. Kim, C.~H. Chu, and S.~M. Shin, ``{ISSAQ: An Integrated Sensing Systems
  for Real-Time Indoor Air Quality Monitoring},'' \emph{IEEE Sensors Journal},
  vol.~14, no.~12, pp. 4230--4244, Dec 2014.

\bibitem{Abdul-Wahab_2015}
S.~A. Abdul-Wahab, S.~C.~F. En, A.~Elkamel, L.~Ahmadi, and K.~Yetilmezsoy, ``A
  review of standards and guidelines set by international bodies for the
  parameters of indoor air quality,'' \emph{Atmospheric Pollution Research},
  vol.~6, no.~5, pp. 751 -- 767, 2015.

\bibitem{Saad_2017}
S.~Saad, A.~Shakaff, A.~Saad, A.~Yusof, A.~Andrew, A.~Zakaria, and A.~Adom,
  ``Development of indoor environmental index: Air quality index and thermal
  comfort index,'' in \emph{AIP Conference Proceedings}, vol. 1808,
  no.~1.\hskip 1em plus 0.5em minus 0.4em\relax AIP Publishing, 2017, p.
  020043.

\bibitem{Wang_2008}
H.~Wang, C.~Tseng, and T.~Hsieh, ``{ Developing an indoor air quality index
  system based on the health risk assessment},'' in \emph{In: Proceedings of
  indoor air 2008}, Copenhagen, Denmark, 2008, paper 749.

\bibitem{Snyder_1995}
J.~W. Snyder, E.~F. Safir, G.~P. Summerville, and R.~A. Middleberg,
  ``Occupational fatality and persistent neurological sequelae after mass
  exposure to hydrogen sulfide,'' \emph{The American Journal of Emergency
  Medicine}, vol.~13, no.~2, pp. 199 -- 203, 1995.

\bibitem{Baelum_1990}
J.~B{\ae}lum, G.~R. Lundgvist, L.~M{\o}lhave, and N.~T. Andersen, ``Human
  response to varying concentrations of toluene,'' \emph{International Archives
  of Occupational and Environmental Health}, vol.~62, no.~1, p.~65, Jan 1990.

\bibitem{VOC_2002}
J.~T. Ayers. ({2002}) { Approaches to a Total (or Grouped) VOC Guideline Final
  Report}.
  \url{https://open.alberta.ca/dataset/8163e248-eed2-41d5-aca7-59d7a7a7b2fc/resource/964e35c4-acb5-4717-a862-adf483cd007b/download/6686approachestoatotalorgroupedvocguideline.pdf}.

\bibitem{national_2011}
NationalToxicologyProgram, ``Ntp 12th report on carcinogens.'' \emph{Report on
  carcinogens: carcinogen profiles}, vol.~12, p. iii, 2011.

\bibitem{OSHA_2018}
\BIBentryALTinterwordspacing
({2018}) {Occupational Safety and Health Administration}.
  \url{https://www.osha.gov}. [Online]. Available: \url{https://www.osha.gov}
\BIBentrySTDinterwordspacing

\bibitem{Huang_2010}
C.-S. Huang, T.~Kawamura, Y.~Toyoda, and A.~Nakao, ``Recent advances in
  hydrogen research as a therapeutic medical gas,'' \emph{Free Radical
  Research}, vol.~44, no.~9, pp. 971--982, 2010.

\bibitem{CCOHS_2018}
\BIBentryALTinterwordspacing
({2018}) {Canadian Centre for Occupational Health and Safety}.
  \url{https://www.ccohs.ca}. [Online]. Available: \url{https://www.ccohs.ca}
\BIBentrySTDinterwordspacing

\bibitem{Hu_2018}
W.~Hu, Y.~Wen, K.~Guan, G.~Jin, and K.~J. Tseng, ``itcm: Toward learning-based
  thermal comfort modeling via pervasive sensing for smart buildings,''
  \emph{IEEE Internet of Things Journal}, vol.~5, no.~5, pp. 4164--4177, Oct
  2018.

\bibitem{Wahid_2013}
Q.~Ha, H.~Wahid, H.~Duc, and M.~Azzi, ``Enhanced radial basis function neural
  networks for ozone level estimation,'' \emph{Neurocomputing}, vol. 155,
  no.~10, pp. 62--70, 2015.

\bibitem{Metia_2018}
S.~Metia, Q.~P. Ha, H.~N. Duc, and M.~Azzi, ``Estimation of power plant
  emissions with unscented kalman filter,'' \emph{IEEE Journal of Selected
  Topics in Applied Earth Observations and Remote Sensing}, vol.~11, no.~8, pp.
  2763--2772, Aug 2018.

\bibitem{Hu_2018a}
X.~{Hu}, H.~{Yuan}, C.~{Zou}, Z.~{Li}, and L.~{Zhang}, ``Co-estimation of state
  of charge and state of health for lithium-ion batteries based on
  fractional-order calculus,'' \emph{IEEE Transactions on Vehicular
  Technology}, vol.~67, no.~11, pp. 10\,319--10\,329, Nov 2018.

\bibitem{Metia_2016}
S.~Metia, S.~D. Oduro, H.~N. Duc, and Q.~Ha, ``Inverse air-pollutant emission
  and prediction using extended fractional kalman filtering,'' \emph{IEEE
  Journal of Selected Topics in Applied Earth Observations and Remote Sensing},
  vol.~9, no.~5, pp. 2051--2063, May 2016.

\bibitem{book:Tepljakov_2017}
A.~Tepljakov, \emph{Fractional-order Modeling and Control of Dynamic Systems
  (Springer Theses)}.\hskip 1em plus 0.5em minus 0.4em\relax Springer, 2017.

\bibitem{book:Ljung_2011}
L.~Ljung, \emph{System Identification: Theory for the User (2nd
  Edition)}.\hskip 1em plus 0.5em minus 0.4em\relax Prentice Hall PTR, Upper
  Saddle River, NJ, USA, 1999.

\end{thebibliography}

\end{document}